\documentclass[14pt,aps,article,pre,groupedaddress,showkeys,showpacs]{revtex4-1}

\usepackage{amsmath,amssymb,bm,graphicx}
\usepackage{mathrsfs}

\usepackage{booktabs}

\usepackage{xcolor}              
\definecolor{a1}{rgb}{0,0,0.8}   
\usepackage{colortbl}	           
\usepackage[colorlinks=true,linktocpage=true,hyperindex=true,pageanchor=true,bookmarks=true]{hyperref} 
\hypersetup{citecolor=a1,linkcolor=a1,menucolor=a1,urlcolor=a1,runcolor=a1}
\hypersetup{citebordercolor=a1,linkbordercolor=a1,urlbordercolor=a1,runbordercolor=a1}
\hypersetup{hypertexnames=false} 
\usepackage{hyperref}

\usepackage{natbib}

\newcommand{\qed}{\hfill \mbox{\raggedright \rule{.07in}{.1in}}}
\usepackage{hyperref}

\begin{document}

\title{Variational Electrodynamics of Atoms}

\author{Jayme De Luca }
\email{jayme.deluca@gmail.com}
\affiliation{Departamento de F\'{i}sica, Universidade Federal de S\~{a}o Carlos, \\ 
Rodovia Washington Luis, km 235, Caixa Postal 676 \\
S\~{a}o Carlos, S\~{a}o Paulo 13565-905 }  
\date{\today }


\begin{abstract}

We generalize Wheeler-Feynman electrodynamics with a variational problem for trajectories that are required to merge continuously into given past and future boundary segments. We prove that the boundary-value problem is well-posed for two classes of boundary data. The well-posed solution in general has velocity discontinuities, henceforth a \emph{broken extremum}. Along regular segments, broken extrema satisfy the Euler-Lagrange \emph{neutral differential delay equations} with state-dependent deviating arguments. At points where velocities are \emph{discontinuous}, broken extrema satisfy the Weierstrass-Erdmann conditions that \emph{energies} and \emph{momenta} are \emph{continuous}. 
Electromagnetic fields of the finite trajectory segments are \emph{derived quantities} that can be extended to a bounded region $\mathcal{B}$ of space-time. Extrema with a finite number $N$ of velocity discontinuities have extended fields defined in $\mathcal{B}$ with the possible exception of $N$ spherical surfaces, and satisfy the integral laws of classical electrodynamics for most surfaces and curves inside $\mathcal{B}$. As an application, we study the hydrogenoid atomic model with mass ratio varying by \emph{three orders of magnitude} to include \emph{hydrogen}, \emph{muonium} and \emph{positronium}. For each model we construct globally bounded trajectories with vanishing far-fields using periodic perturbations of circular orbits. Our model uses solutions of the neutral differential delay equations along regular segments and a variational approximation for the head-on collisional segments. Each hydrogenoid model predicts a discrete set of finitely measured neighbourhoods of periodic orbits with vanishing far-fields right at the correct atomic magnitude and in quantitative and qualitative agreement with experiment and quantum mechanics. The spacings between consecutive discrete angular momenta agree with Planck's constant within thirty-percent, while orbital frequencies agree with a corresponding spectroscopic line within a few percent.
\end{abstract}
\maketitle



\section{Introduction}
\label{introduction}
 
We generalize Wheeler-Feynman electrodynamics \cite{Whe-Fey} with a variational principle whose extrema are required to satisfy a boundary-value problem \cite{JMP2009, Minimizers}. As in all variational problems, there is no guarantee that smooth classical solutions exist. In fact, here we prove that one should expect \emph{piecewise smooth} extrema. For generic boundary data, solutions are continuous trajectories with velocity discontinuity points, henceforth corner points \cite{Gelfand}.
\par
Piecewise smooth extrema satisfy the Wheeler-Feynman \emph{neutral differential delay} equations with \emph{state-dependent} deviating arguments along smooth segments \cite{JMP2009, Minimizers}. At corner points piecewise smooth extrema satisfy the Weierstrass-Erdmann conditions that \emph{partial energies} and \emph{momenta} are \emph{continuous} \cite{Gelfand}.
\par

For two special classes of boundary data we prove that the variational principle is well-posed, i.e., there exists a unique solution depending continuously on the boundary data. The piecewise smooth solutions define generalized electromagnetic fields inside a bounded region $\mathcal{B}$ of space-time by extension. We show that the extended fields satisfy the integral laws of classical electrodynamics inside $\mathcal{B}$, i.e., Gauss's surface integral law for the electric field, Gauss's surface integral law for the magnetic field, Ampere's law and Faraday's induction law in \emph{integral} form \cite{Jackson}.
\par

Wheeler and Feynman derived \emph{neutral differential delay equations} (NDDE) for the motion of point charges \cite{Whe-Fey}. NDDE are functional differential equations whose qualitative behaviour has just begun to be understood \cite{JackHale,BellenZennaro,HKWW06}. In qualitative agreement with the existence of broken extrema, solutions of NDDE must be defined piecewise \cite{Minimizers}. Solution continuation leaves a set of points where trajectories are \emph {not} differentiable \cite{BellenZennaro} (for a pedestrian explanation see Appendix A of Ref. \cite{double-slit}). In numerical analysis, a velocity discontinuity point is called a breaking point \cite{BellenZennaro}, while in variational calculus (and here) the name corner point is used \cite{Gelfand}. 
\par

Surprizingly, atomic models become \emph{sensible} in variational electrodynamics. More specifically, our generalized electrodynamics allows globally bounded two-body orbits with vanishing far-fields, thus introducing bounded motions along which an atom is isolated from disturbing/being disturbed by other atoms. The essential ingredient is precisely corner points. It is proved in Ref. \cite{Minimizers} that globally bounded two-body orbits with vanishing far-fields \emph{must} have corner points. 
\par
We attempt to validate our theory by exploring the hydrogenoid atomic model with mass-ratio varying by \emph{three orders of magnitude} to include the hydrogen, muonium and positronium atoms. Our model constructs periodic orbits with vanishing far-fields and having regular segments where the NDDE and deviating arguments are linearized, while on the (thin) boundary-layer segments a variational approximation is used.  
\par
In the three cases of hydrogen, muonium and positronium,  a discrete set of finitely measured neighbourhoods of orbits with vanishing far-fields have frequencies in agreement with Quantum Mechanics (QM), within a few percent. The qualitative agreements with QM are (i) the angular momenta of the unperturbed circular orbits are approximately
 integer multiples of a basic angular momentum agreeing with Planck's constant within thirty percent;
(ii) the emitted frequency is the difference
of two eigenvalues of a suitable linear problem; and (iii) the Weierstrass-Erdmann conditions involve the continuity of \emph{momenta} and \emph{energies}, which
are the relevant quantities of QM. 

\par
This paper is long and the reader can separate it in two parts, as follows. Sections  (\ref{Section II}-\ref{Section IV}) outline the theory of variational electrodynamics: Section (\ref{Section II}) introduces the boundary-value problem, the variational structure, and the Weierstrass-Erdmann conditions. In Section \ref{Section III} we prove that the boundary-value problem is well-posed. Section \ref{Section IV} explains variational electrodynamics as an extension of Wheeler-Feynman electrodynamics by discussing the conditions for the validity of the integral laws, which are the experimental basis of classical electrodynamics. Section \ref{Section IV} also discusses invariant manifolds and a generalized absorber condition. 
In Sections (\ref{Section V}-\ref{Section VIII}) we study atomic models using the theory of Sections  (\ref{Section II}-\ref{Section IV}); Section \ref{Section V} introduces the circular orbits and magnitudes in the limit of small delay angles. In Section \ref{Section VI} we linearize the Wheeler-Feynman NDDE about circular orbits and explain the infinite number of linearly unstable transversal modes. Section \ref{Section VII} discusses the boundary-layer theory and application of the
Weierstrass-Erdmann conditions. In Section \ref{Section VIII} we validate our theory by comparing the predictions of the hydrogenoid model with the experimental magnitudes of \emph{hydrogen}, \emph{muonium} and \emph{positronium}. 
Last, in Section \ref{Section IX} we put the discussions and conclusion.

\section{Boundary-value problem}
\label{Section II}

We henceforth use a unit system where the speed of light is $c \equiv 1$ and the electronic charge and electronic mass are respectively $e_{1} \equiv -1$ and $m_{1} \equiv 1$. The protonic charge and protonic mass in our unit system turn out to be respectively $e_{2}= 1$ and $m_{2}=1836.1526$. 
\par
 A seemingly essential ingredient for a viable physical (and mathematical) theory is the minimization of a suitably defined functional, henceforth a \emph{variational principle}. A useful paradigm is the principle of least action of classical mechanics specialized to the Kepler two-body problem \cite{Feynman_Lectures}. Hamilton's principle states that the action functional assumes an extremum on the classical two-body orbit of a finite time-interval, when considered in the class of $C^2$ smooth trajectories sharing the same endpoints.
\par
The principle of least action \cite{Feynman_Lectures} defines a two-point boundary problem for the ordinary differential equations (ODE) of classical mechanics \cite{Fox,Petzold}, often called a \emph{shooting problem} to distinguish from the \emph{initial value problem} \cite{Fox,Petzold}. Motivated by Wheeler-Feynman electrodynamics, Ref. \cite{JMP2009} constructed a Poincar\'{e}-invariant action principle at the expense of introducing the unusual boundary conditions explained below.

\par
Physical trajectories should have a velocity lesser than the speed of light, henceforth sub-luminal trajectories. We describe our relativistic trajectories in the usual Minkowski space, where every point $ P \equiv (t, \mathbf{x})$ has a time and a Cartesian coordinate \cite{JMP2009,JLMartin}. A point $ P_{+} \equiv (t_{j+}, \mathbf{x}_{j+})$ belongs to the \emph{future of $P$} when 
$ t_{j+} > t + \Vert {\mathbf{x}}_{j+}-{\mathbf{x}} \Vert$, while a point $ P_{-} \equiv (t_{j-}, \mathbf{x}_{j-})$ belongs to the \emph{past of $P$} when $ t_{j-} < t - \Vert {\mathbf{x}}_{j-}-{\mathbf{x}} \Vert$ \cite{JLMartin}. 
The set of points neither in the future of $P$ nor in the past of $P$ is defined as the \emph{elsewhere of $P$} \cite{JLMartin}.  A point $P_{\pm}$ along trajectory $\mathbf{x}_{j} \equiv \mathbf{x}_{j}(t_j)$ is in the \emph{light-cone} relation with another point $P$ if
\begin{equation}
t_{j\pm } = t \pm \Vert {\mathbf{x}}_{j \pm}(t_{j\pm})-{\mathbf{x}} \Vert.
\label{light-cone}
\end{equation}
Equation (\ref{light-cone}) is an implicit state-dependency on $ \mathbf{x}_{j}(t_{j\pm})$, henceforth the light-cone condition or the Einstein locality condition. In Eq. (\ref{light-cone}), the plus sign defines the future light-cone of $P$ and the minus sign defines the past light-cone of $P$.  
 \par 
To continue trajectory $1$ from an initial point $O_{1} \equiv (t_{O_1},\mathbf{x}_1(t_{O_1}) )$, a \emph{relativistic} variational principle needs the whole intersection of trajectory $2$ with the \emph{elsewhere} of $O_1$, which is a \emph{finite segment} of trajectory $2$. For the classical principle of least action, the elsewhere of $O_1$ degenerates into the initial point of trajectory $2$. Last, at the end-point $L_2$ of trajectory $2$, the relativistic least-action principle needs the intersection of trajectory $1$ with the elsewhere of $L_2$, again a finite segment of trajectory $1$ rather than a simple endpoint. 
\par
\begin{figure}[h!]
   \centering
    \includegraphics[scale=0.45]{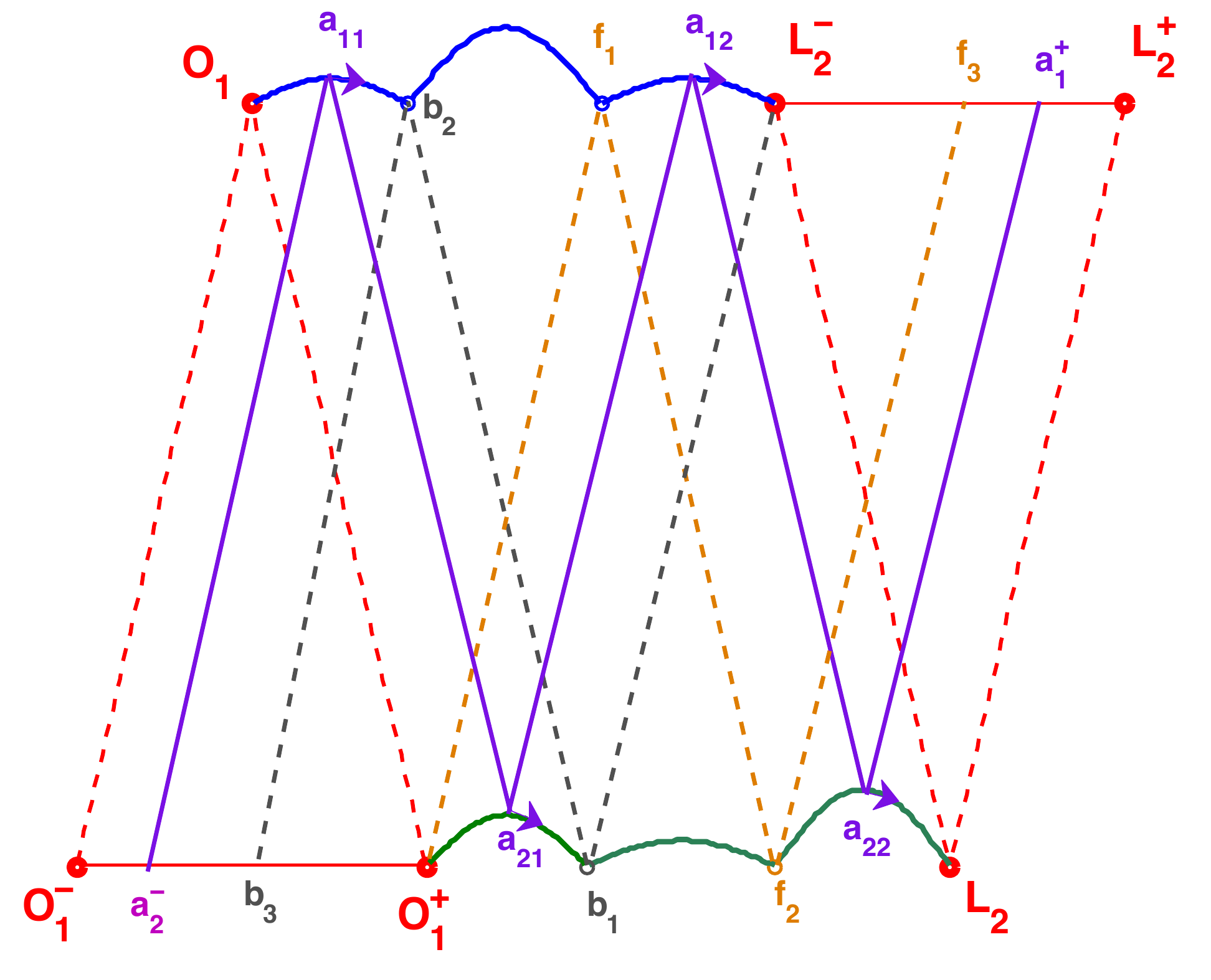} 
   \caption{Schematic illustration of the boundaries in $\mathbb{R}^4$, i.e., (a)
initial point $O_{1} \equiv (t_{O_1},\mathbf{x}_1(t_{O_1}) )$ of trajectory  $1$
and the respective elsewhere boundary segment of $\mathbf{x}_2(t_2)$ for $ t_2\in [t_{O_1^-},t_{O_1^+}]$ (solid red line); (b)  endpoint $L_{2} \equiv (t_{L_2}, \mathbf{x}
_2(t_{L_2}))$ of trajectory $2$ and respective elsewhere boundary segment of $%
\mathbf{x}_1(t_1)$ for $t_1\in[t_{L_2^-},t_{L_2^+}]$ (solid red line). Trajectories $\mathbf{x}_1(t_1)$
 for $t_1\in[t_{O_1},t_{L_2^-}]$ (solid blue line) and $\mathbf{x}_2(t_2)$ for $t_2\in[%
t_{O_1^+},t_{L_2}]$ (solid green line) are determined by the extremum condition. The principal sewing chains are also illustrated; i.e., the forward sewing chain of $O^+_1$, $(O^+_1,f_1,f_2,f_3)$, (broken golden line) and the backwards sewing chain of $L_2^-$, $(L_2^-, b_1,b_2,b_3)$ (broken dark line). Sewing chain $(a^{-}_2, a_{11},a_{21},a_{12},a_{22},a^{+}_1)$ has two points on each trajectory (solid violet line). Violet arrows are directions of integration to be explained below. Arbitrary units.} 
\label{Squema}
\end{figure}

The unusual boundary conditions for a relativistic variational principle are illustrated in FIG. \ref{Squema}, i.e., (a) the initial point $O_1$ of
trajectory $1$ and the respective boundary-segment of trajectory $2$ inside
the light-cone of $O_1$ (red triangle of FIG. \ref{Squema}), and (b) the final point $L_2$ of 
trajectory $2$ and the corresponding boundary-segment of trajectory
$1$ inside the light-cone of $L_2$ (upside down red triangle of FIG. \ref{Squema}).

\par
 Along continuous and piecewise $C^1$ \emph{sub-luminal} trajectories satisfying the boundaries of FIG. \ref{Squema}, the past and the future light-cone conditions (\ref{light-cone}) have \emph{unique solutions} $ t_{j\pm}(t, \mathbf{x})$ \cite{CAM}. Illustrated in FIG. \ref{Squema} are also the forward light-cone rays starting from $O^+_1$ and moving with the future light-cone condition (\ref{light-cone}) to $f_1$, $f_2$ and $f_3$, and the backwards light-cone rays starting from $L^-_2$ and moving with the past light-cone condition (\ref{light-cone}) to $b_1$, $b_2$ and $b_3$, henceforth called the forward and backwards principal sewing chains, respectively. 
\par
The action functional is a sum of four integrals: two  \textit{local} integrals each involving one trajectory, $\int T_i (\mathbf{x}_i,\mathbf{\dot{x}}_i)dt_i$; and two \textit{interaction} integrals depending on both positions and velocities, where one position/velocity is evaluated at a deviating time argument, $\int V^\pm_{ij}(\mathbf{x}_i, \mathbf{\dot{x}}_i,\mathbf{x}_{j\pm},\mathbf{\dot{x}}_{j\pm})dt_i$. The action functional can be expressed in two equivalent forms, i.e.,
\begin{align} \label{aFokker}
S \equiv &\negthickspace \int_{t_{O_1^{+}}}^{t_{L_{2}}}\negthickspace T_{2}dt_{2}
+\negthickspace \int_{t_{O_{1}}}^{t_{L_2^{-}}}\negthickspace  T_{1}dt_{1}+\underbrace{ \int_{t_{O_{1}}}^{t_{L_2^{+}}} \negthickspace V_{12}^-dt_{1}}
+ \underbrace{\negthickspace \int_{t_{O_1}}^{t_{L_2^-}}\negthickspace  V_{12}^+dt_{1}},\\
& \qquad \qquad   \qquad\qquad\qquad\qquad\;  \; \;  \;\parallel \qquad\qquad\quad\; \; \parallel \notag\\
\negthickspace \negthickspace =&\negthickspace \int_{t_{O_{1}}}^{t_{L_2^{-}}} \negthickspace T_{1}dt_{1}
+\negthickspace \int_{t_{O_1^{+}}}^{t_{L_{2}}}\negthickspace \negthickspace T_{2}dt_{2}
+ \overbrace{ \int_{t_{O_1^-}}^{t_{L_2}} \negthickspace V_{21}^+dt_{2}}
+ \overbrace{ \int_{t_{O_1^+}}^{t_{L_{2}}} \negthickspace V_{21}^-dt_{2}},
\label{L2}
\end{align}
where vertical braces under each interaction integral indicate equivalence by a change of the integration variable. The Jacobian for each change of variable from $t_i$ to $t_{j\pm}(t_i)$ is equal to the derivative of the delayed time $t_{j\pm}(t_{i}, \mathbf{x}_j(t_i))$
evaluated along the orbit, as obtained taking a derivative of the implicit condition (\ref{light-cone}) with $\mathbf{x} \equiv \mathbf{x}_i(t_i)$, 
\begin{equation}
\frac{d t_{j\pm }}{d t_{i}}= \frac{(1 \pm \mathbf{n}_{j\pm} \cdot \mathbf{v}_i)}{(1 \pm \mathbf{n}_{j\pm} \cdot \mathbf{v}_{j\pm})},  \label{t2difft1}
\end{equation}
and explained in Refs. \cite{JMP2009,CAM}. One can thus express the interaction terms by either integrals over $t_1$ (Eq. (\ref{aFokker})) or by integrals over $t_2$ (Eq. (\ref{L2})).   
In principle, an arbitrary variational structure could be defined using generic $V's$ on line (\ref{aFokker}), which would in turn determine the $V's $ on line (\ref{L2}) by changing variables with (\ref{t2difft1}) or the equivalent Jacobian if the constraints were other than the light-cone conditions (\ref{light-cone}). 
\par
Here we consider only the variational structure defined by constraints (\ref{light-cone}) and functionals (\ref{aFokker}) and (\ref{L2}) with
\begin{eqnarray}
T_i\equiv m_i ( 1-\sqrt{1-\mathbf{v}_i^2 }   \;),
\qquad \\
V^\pm_{ij}(\mathbf{x}_i,\mathbf{\dot{x}}_i,\mathbf{x}_{j\pm},\mathbf{\dot{x}}_{j\pm})
\equiv -\frac{e_i e_j (1-\mathbf{v}_i \cdot \mathbf{v}_{j\pm}) }{2r_{j\pm}(1\pm \mathbf{n}_{j\pm}\cdot \mathbf{v}_{j\pm})  },  \label{EU}
\end{eqnarray}
where $j \equiv 3-i$ and $i=1,2$; henceforth variational electrodynamics. For the hydrogenoid problem we henceforth replace $e_i e_j \equiv -1$ and carry an arbitrary protonic mass, for which case $V^{\pm}_{12}  \geq 0$ and $V^{\pm}_{21}  \geq 0$, thus defining a semi-bounded action functional (\ref{aFokker}) ($S \geq 0$). 
\par
The variational problem is to find the trajectory segments $(O_{1}, L_2^-)$ (blue) and $(O_1^+, L_2)$ (green) between the endpoints of FIG. \ref{Squema}.  For the linear variation, trajectory $2$ is to be kept fixed while trajectory $1$ is varied, and line (\ref{aFokker}) with the first term kept constant defines partial Lagrangian $1$.
Vice-versa, trajectory $1$ is to be kept fixed while trajectory $2$ is varied, and line (\ref{L2}) with the first term frozen defines partial Lagrangian $2$.

\par
Next we discuss acceptable trajectories for the variational problem. The classical calculus of variations requires at least a neighbourhood in a normed space of piecewise-smooth continuous trajectories \cite{Gelfand}. Specific difficulties are (i) along acceptable trajectories satisfying the boundaries of FIG. \ref{Squema}, functionals (\ref{aFokker}) and (\ref{L2}) require existence of unique advanced/retarded arguments $t_{2\pm}(t_1)$, $\forall \, t_1 \in [t_{O_1}, t_{L^+ _2} ]$ and $t_{1\pm}(t_2)$, $\forall \, t_2 \in [t_{O_1^-}, t_{L_2} ]$; and (ii) functional-analytic results require a whole \emph{domain} in which functionals (\ref{aFokker}) and (\ref{L2}) are well defined, e.g., a normed linear space. 
\par
To satisfy (i) a neighbourhood of $C^1$ smooth sub-luminal trajectories suffices, as guaranteed by Lemma 1 of Ref. \cite{CAM}. We remark that Lemma 1 of \cite{CAM} can be extended to continuous trajectories that are sub-luminal \emph{almost everywhere}, a measure-theoretic extension not pursued here. As regards (ii) we notice that the integrands of functionals (\ref{aFokker}) and (\ref{L2}) include denominators that should be non-zero outside sets of zero measure, as discussed in Ref. \cite{Gordon2} for the Kepler problem. 
\par
To study critical points using the modern topological theorems \cite{Brezis, Jabri} would require a reflexive space (Banach or Hilbert) where (\ref{aFokker}) and (\ref{L2}) are finitely integrable \cite{Gordon2} and Frech\'{e}t differentiable \cite{Jabri}. Such ambitious goal is beyond the present work. Henceforth we study functional minimization restricted to neighbourhoods of non-collisional sub-luminal trajectories, along which the denominators of (\ref{aFokker}) and (\ref{L2}) are everywhere finite. The topology used is that of the normed space of continuous and piecewise $C^2$ functions, henceforth $\widehat{C}^2$.
\par
For $C^2$ smooth extrema, the critical point conditions are the Euler-Lagrange equations of the integrands of Eqs. (\ref{aFokker}) and (\ref{L2}) with the respective first term dropped, henceforth the partial Lagrangians defined by
\begin{equation}
\mathscr{L}_i \equiv T_i -\mathbf{v}_i \cdot \mathbf{A}_j + U_j, \label{partial}
\end{equation}
 where
\begin{eqnarray}
  \negthickspace \negthickspace  \mathbf{A}_j & \equiv &  \frac{\mathbf{{v}}_{j-} }{2r_{j-}(1-\mathbf{n}_{j-}\cdot \mathbf{v}_{j-})} +
\frac{\mathbf{{v}}_{j+}}{2r_{j+}(1+\mathbf{n}_{j+}\cdot \mathbf{v}_{j+})}, \notag \\
\negthickspace  \negthickspace \negthickspace \negthickspace {U}_j & \equiv & \frac{1 }{2r_{j-}(1-\mathbf{n}_{j-}\cdot \mathbf{v}_{j-})} +
\frac{1}{2r_{j+}(1+\mathbf{n}_{j+}\cdot \mathbf{v}_{j+})}, \notag
\end{eqnarray}
for $j \equiv 3-i$ and $i=1,2$. In Eq. (\ref{partial}), $\mathbf{A}_j$ is the vector potential of particle $j$.
\par
The Euler-Lagrange equation of the above defined $\mathscr{L}_i$ yields the Lorentz-force law (the equation of motion of Wheeler and Feynman)
\begin{equation}
m_{i}\frac{d}{dt}\left(\frac{\mathbf{v}_{i
}}{\sqrt{1-\mathbf{v}_{i}^{2}}}\right)= e_i (\mathbf{E}_j(t,\mathbf{x}_i)+ \mathbf{v}_{i
} \times \mathbf{B}_j(t,\mathbf{x}_i)) \label{Lorentz}, 
\end{equation}
for $i=1,2$ \cite{Whe-Fey}. In Eq. (\ref{Lorentz}), the Li\'{e}nard-Wiechert fields of the other charge ($j \equiv 3-i$) are 
\begin{eqnarray}
 \mathbf{E}_{j}(t,\mathbf{x})& \equiv &  \frac{1}{2}(\mathbf{E}_{j+} + \mathbf{E}_{j-}), \label{sumE} \\
  \mathbf{B}_{j}(t,\mathbf{x})& \equiv &  \frac{1}{2}(\mathbf{B}_{j+} + \mathbf{B}_{j-}),\label{sumB}
 \end{eqnarray}
with
 \begin{eqnarray}
 \mathbf{E}_{j\pm}(t,\mathbf{x}) &  \equiv &e_j \{\frac{\mathbf{u}_{j\pm} } {\gamma_{j\pm}^2 r^{2}_{j\pm}}+\frac{\mathbf{n}_{j\pm}\times (\mathbf{u}_{j\pm} \times \mathbf{a}_{j\pm} )} { r_{j\pm}} \}, 
\label{electric} \\
\mathbf{u}_{j\pm} (t,\mathbf{x})& \equiv & \frac {( \mathbf{n}_{j\pm }\pm \mathbf{v}_{j\pm} )}{(1\pm \mathbf{n}_{j\pm}\cdot \mathbf{v}_{j\pm})^{3}}, \label{defu}\\ 
 \mathbf{B}_{j\pm}(t,\mathbf{x})& \equiv &\mp \mathbf{n}_{j\pm}\times \mathbf{E}_{j\pm},
\label{magnetic}
\end{eqnarray}
where $\gamma_{j\pm} \equiv (1-\mathbf{v}^2_{j\pm})^{-1/2}$, and $\mathbf{v}_{j\pm} \equiv d \mathbf{x}_j/dt|_{t=t_{j\pm}}$ and $\mathbf{a}_{j\pm} \equiv d^2 \mathbf{x}_j/dt^2 |_{t=t_{j\pm}}$ are the velocity and acceleration of charge $j$ evaluated at the advanced/retarded times $t_{j\pm}$ defined by Eq. (\ref{light-cone}).  Last, in Eq. (\ref{electric}) the distance in light-cone is a scalar function of $(t,\mathbf{x})$ defined by
\begin{eqnarray}
 r_{j\pm}\equiv  \Vert {{\mathbf{x} -\mathbf{x}}_{j}(t_{j\pm})} \Vert , \label{fieldrj}
\end{eqnarray}
 and 
 \begin{eqnarray}
 \mathbf{n}_{j\pm } \equiv  (\mathbf{x} -\mathbf{x}_{j} (t_{j\pm})) / r_{j\pm}, \label{defn}  
 \end{eqnarray}
 is a unit vector from position $\mathbf{x}_{j}(t_{j\pm})$ to point $\mathbf{x}$ \cite{Whe-Fey, Jackson}. Equation (\ref{Lorentz}) is a neutral differential delay equation (NDDE) with state-dependent deviating arguments \cite{Whe-Fey, Minimizers}.
 \par
In the following we study the wider class of \emph{piecewise smooth} continuous extrema having a finite number of corner points, henceforth  \textit{broken extrema} \cite{Gelfand}. Piecewise smooth extrema have the following nice properties: (i) inside intervals where trajectory and deviating arguments are $C^2$ smooth, broken extrema satisfy the Euler-Lagrange equations (\ref{Lorentz}); and (ii) at corner points extremal trajectories satisfy the Weierstrass-Erdmann conditions explained below \cite{Gelfand}. 

\par
The first Weierstrass-Erdmann condition \cite{Gelfand} is the continuity of the momentum of partial Lagrangian $i$ (as defined by Eq. (\ref{partial})), i.e.,
\begin{eqnarray}
\frac{\partial \mathscr{L}_i }{ \partial \mathbf{{v}}_{i}} = \frac{m_{i} \mathbf{{v}}_{i}  }{ \sqrt{1 - \mathbf{v}_i^2}  }   
-\frac{\mathbf{{v}}_{j-} }{2r_{j-}(1-\mathbf{n}_{j-}\cdot \mathbf{v}_{j-})} \notag \\ -
\frac{\mathbf{{v}}_{j+}}{2r_{j+}(1+\mathbf{n}_{j+}\cdot \mathbf{v}_{j+})},
\label{momentum1} \\
= \frac{m_{i} \mathbf{{v}}_{i}  }{ \sqrt{1 - \mathbf{v}_i^2}  }   -{\mathbf{A}_j}. \label{preWE1}
\end{eqnarray}
  Notice that Eq.  (\ref{momentum1}) includes the past/future velocities of charge $j \equiv 3-i$. Should the extremal trajectory of particle $i$ have a velocity discontinuity at time $t_i$, the trajectory of particle $j$ must compensate with a corner point in light-cone at either $t_{j-}$ or $t_{j+}$, in order to make the right-hand-side of (\ref{momentum1}) continuous. 
 \par
  Here we use the name \emph{partial energy} to distinguish from the constant value of the Hamiltonian along a Hamiltonian dynamics. After the no-interaction theorem \cite{Currie, Marmo} we know that a finite-dimensional Hamiltonian does not exist for the electromagnetic two-body problem, even though partial energies are introduced below by an energy-looking formula.
\par
 The partial energy of partial Lagrangian (\ref{partial}) is defined by
\begin{equation}
E_i \equiv \mathbf{v}_i \cdot \frac{\partial \mathscr{L}_i }{ \partial \mathbf{{v}}_{i}} -\mathscr{L}_i = \frac {m_i} {\sqrt{1-\mathbf{v}_i^2}}  -  U_j  \label{WE2}.
\end{equation}
The second Weierstrass-Erdmann condition \cite{Gelfand} is that the partial energy (\ref{WE2}) is \emph{continuous} across each corner point. In Eq. (\ref{WE2}) index $j$ is defined by the usual $j\equiv 3-i$, e.g. for $i=2$ we have $j=1$. We stress that the $E_i$ in Eq. (\ref{WE2}) are \emph{not} constants of the motion. The partial energy is a property of each particular corner (perhaps different for different corners). Each $E_i$ is conserved only \emph{across} a particular corner in the sense of having the same value to the left and to the right of \emph{that} corner.
\par
To express the vanishing of the momentum jump (\ref{preWE1}) across a corner point we introduce an upper index $l$ or $r$ to indicate respectively left-velocity or right-velocity at the breaking point. Using Eq. (\ref{WE2}) to eliminate the mechanical momentum from Eq. (\ref{momentum1}), we obtain the combined \emph{necessary condition} for a corner point, 
\begin{eqnarray}
  \Delta (\frac{\partial \mathscr{L}_i }{ \partial \mathbf{{v}}_{i}}) = E_i \Delta \mathbf{v}_{i}    \notag \\
+\left( \frac{\Delta \mathbf{{v}}_{i}-(\mathbf{n}_{j-}\cdot \mathbf{v}_{j-}^l ) \mathbf{v}_{i}^r  +(\mathbf{n}_{j-}\cdot \mathbf{v}_{j-}^r ) \mathbf{v}_{i}^l   }{2r_{j-}(1-\mathbf{n}_{j-}\cdot \mathbf{v}_{j-}^l)(1-\mathbf{n}_{j-}\cdot \mathbf{v}_{j-}^r)} \right) \notag \\ +\left(
\frac{\Delta \mathbf{{v}}_{i} +(\mathbf{n}_{j+}\cdot \mathbf{v}_{j+}^l ) \mathbf{v}_{i}^r  -(\mathbf{n}_{j+}\cdot \mathbf{v}_{j+}^r ) \mathbf{v}_i^l }{2r_{j+}(1+\mathbf{n}_{j+}\cdot \mathbf{v}_{j+}^{l})(1+\mathbf{n}_{j+}\cdot \mathbf{v}_{j+}^{r})} \right)
\notag \\
-\left( \frac{\Delta \mathbf{{v}}_{j-}-\mathbf{n}_{j-}\times(\mathbf{v}_{j-}^r \times \mathbf{v}_{j-}^l )  }{2r_{j-}(1-\mathbf{n}_{j-}\cdot \mathbf{v}_{j-}^l)(1-\mathbf{n}_{j-}\cdot \mathbf{v}_{j-}^r)} \right) \notag \\ -\left(
\frac{\Delta \mathbf{{v}}_{j+}+\mathbf{n}_{j+}\times(\mathbf{v}_{j+}^r \times \mathbf{v}_{j+}^l)}{2r_{j+}(1+\mathbf{n}_{j+}\cdot \mathbf{v}_{j+}^{l})(1+\mathbf{n}_{j+}\cdot \mathbf{v}_{j+}^{r})} \right)
\notag \\ \equiv 0, \label{WE1}
\end{eqnarray}
where $\Delta \mathbf{{v}}_{i} \equiv   \mathbf{{v}}_{i}^r- \mathbf{{v}}_{i}^l$ , \;  $\Delta \mathbf{{v}}_{j\pm} \equiv   \mathbf{{v}}_{j\pm}^r- \mathbf{{v}}_{j\pm}^l$ and $j \equiv 3-i$. Equation (\ref{WE1}) is a \emph{nonlinear} condition for the jumping velocities, a necessary condition involving the $E_i$ to be adjusted such that (\ref{WE2}) is continuous at that corner. In Section \ref{Section III}, Eq. (\ref{WE1}) is linearized for small jumps by replacing $ \mathbf{{v}}_{i}^r \rightarrow \mathbf{{v}}_{i}^l +\Delta \mathbf{{v}}_{i} $ and $ \mathbf{{v}}_{j\pm}^r \rightarrow  \mathbf{{v}}_{j\pm}^l  + \Delta \mathbf{{v}}_{j\pm}$ and expanding up to linear order on the $\Delta \mathbf{v}_i\,$, in which case the partial energies appear as eigenvalues of the linearized problem (\ref{WE1}). In Section \ref{Section VII}, the fully nonlinear condition (\ref{WE1}) is used.

\section{Well-posedness}
\label{Section III}

\begin{figure}[h!]
   \centering
  \includegraphics[scale=0.45]{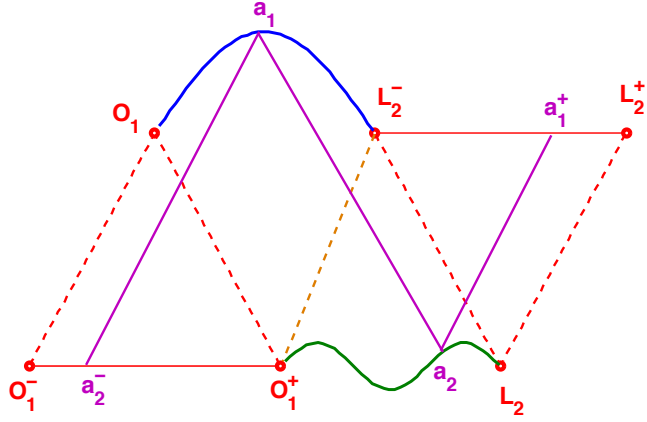} 
\caption{ Schematic illustration of the point-plus-elsewhere boundary sets in $\mathbb{R}^4$ for a boundary-value problem of shortest length, i.e., (a)
initial point $O_{1} \equiv (t_{O_1},\mathbf{x}_1(t_{O_1}) )$ of trajectory  $1$
and respective elsewhere boundary segment of $\mathbf{x}_2(t_2)$ for $ t_2\in [t_{O_1^-},t_{O_1^+}]$ (solid red line); and (b)  endpoint $L_{2} \equiv (t_{L_2}, \mathbf{x}
_2(t_{L_2}))$ of trajectory $2$ and respective elsewhere boundary segment $%
\mathbf{x}_1(t_1)$ for $t_1\in[t_{L_2^-},t_{L_2^+}]$ (solid red line). Trajectories $\mathbf{x}_1(t_1)$
 for $t_1\in[t_{O_1},t_{L_2^-}]$ (solid blue line) and $\mathbf{x}_2(t_2)$ for $t_2\in[%
t_{O_1^+},t_{L_2}]$ (solid green line) are determined by the extremum condition. Sewing chain ($\mathbf{a}_2^- $, $ \mathbf{a}_1$ , $\mathbf{a}_2$, $\mathbf{a}_1^+$) starts from $\mathbf{a}_1$ along trajectory $1$ and moves (forward and backwards) until the boundary segments (solid violet line). }
  \label{fig8}
\end{figure}

As mentioned in \cite{JMP2009,CAM} and illustrated in FIG. \ref{fig8}, the shortest-length boundary-value problem is when $L_2^-$ is in the forward light-cone of $O_1^+$. Otherwise the supposedly independent past and future histories interact in light-cone, an absurdity. 
\par
In the following we prove that the boundary-value problem is well-posed for $C^2$ boundary segments of the above defined shortest length.  We further assume boundary segments sufficiently close to segments of circular orbits of small-delay-angles \cite{Schild}, with continuity defined with the $\widehat{C}^2$ topology of continuous and piecewise smooth trajectories.

  It is instructive to translate the schematics of FIG. \ref{Squema} and FIG. \ref{fig8} in terms of the magnitudes of circular orbits with small delay angles, whose light-cone-distance is $r_b \equiv 1/ \mu \theta^2$, (Eq. (\ref{RB})), and velocity $| \mathbf{v}| \simeq \theta \ll 1$, (Eq. (\ref{defvelocity})). The time span of each circular trajectory of FIG. \ref{fig8} is $\Delta t_j \equiv  t_{{L}^{-}_2}  - t_{{O}_1}=t_{{O}^{+}_1}  - t_{{L}_2}=2 r_b$, for $j=1,2$ (blue and green segments). The shortest time of flight is two light-cone distances $r_b$, and for small $\theta$ these are almost straight-line constant-velocity trajectories (FIG. \ref{fig1} with small $\theta$). 
  \par
  The two-point-boundary-value problem with $\Delta t_j =2r_b $ is related to the initial value problem by a linear one-to-one map, i.e.,  $\mathbf{x}(\Delta t)= \mathbf{x}(0) + \int_0^{\Delta t} \mathbf{v}(t')  dt \simeq  \mathbf{x}(0)+2r_b \mathbf{v}(0)$ \cite{Petzold}. Below we are dealing with small perturbations of the former map, in which case the implicit function theorem allows one to control the end-point by adjusting the initial velocity.
\par
\textbf{Theorem I}:  (i) For $C^2$ boundaries of shortest-length (FIG. \ref{fig8}), the \emph{unique} solution depends \emph{continuously} on boundary data that are sufficiently close to segments of a small-delay-angle circular orbit (again, continuity in the $\widehat{C}^2$ topology). (ii) Generically, the velocities are discontinuous at $O_1^+$ and $L^{-}_2$.
\par
\textbf{Proof}: (i) As illustrated in FIG. \ref{fig8}, for the shortest-length case the past light-cone of $\mathbf{a}_1$ falls on the past history segment,  i.e., point $\mathbf{a}^-_2$ of FIG. \ref{fig8}. Likewise, the future light-cone of $\mathbf{a}_2$ is on the future history segment (illustrated by $\mathbf{a}^+_1$ in FIG. \ref{fig8}). Next we write the equations for accelerations $\mathbf{a}_1$ and $\mathbf{a}_2$, which interact in light-cone.  The equations of motion (\ref{Hans}) yield
\begin{eqnarray}
 \frac{m_1 \mathbf{a}_{1
}}{\sqrt{1-\mathbf{v}_{1}^{2}}}& =& \mathbb{A} \mathbf{a}_{2} + \mathbf{F}^{-}_2(\mathbf{x}_1,\mathbf{x}_{2+}, \mathbf{v}_1,\mathbf{v}_{2+}),   \label{short1}\\
\frac{m_2 \mathbf{a}_{2
}}{\sqrt{1-\mathbf{v}_{2+}^{2}}}  &=& \mathbb{B} \mathbf{a}_{1} +\mathbf{F}^+_{1}(\mathbf{x}_{2+},\mathbf{x}_{1} \mathbf{v}_{2+},\mathbf{v}_{1}).  \; 
\label{short2}
\end{eqnarray} 
\par
On the right-hand-sides of Eqs. (\ref{short1}) and (\ref{short2}) we separated the linear dependence on the other particle's running acceleration across the light-cone.
In Eq. (\ref{short1}), vector $\mathbf{F}^{-}_2$ depends \emph{continuously} on the past-history segment's position, velocity and acceleration. Analogously, in Eq. (\ref{short2}), $\mathbf{F}^+_1$ depends continuously on the future history segment's position, velocity and acceleration (again, continuity with the $\widehat{C}^2$ topology). 
\par
Eliminating $\mathbf{a}_2$ from the right-hand{\color{blue}-}side of Eq. (\ref{short1}) with Eq. (\ref{short2}), and eliminating $\mathbf{a}_1$ from the right-hand-side of Eq. (\ref{short2}) with Eq. (\ref{short1}), yields
\begin{eqnarray}
 (m_1 \mathbb{I}_3 - \frac{1}{m_2} \mathbb{\tilde{A}}\mathbb{\tilde{B}} )\mathbf{a}_{1}& = \mathbf{\tilde F}_1(\mathbf{x}_1,\mathbf{x}_{2+}, \mathbf{v}_1,\mathbf{v}_{2+}),   \label{ort1}\\
(m_2 \mathbb{I}_3 - \frac{1}{m_1 } \mathbb{\tilde{B}}\mathbb{\tilde{A}})\mathbf{a}_{2}& = \mathbf{\tilde F}_{2}(\mathbf{x}_1,\mathbf{x}_{2+}, \mathbf{v}_1,\mathbf{v}_{2+}),   \; \; \; \; \label{ort2}
\end{eqnarray} 
where $\tilde{\mathbb{A}} \equiv \sqrt{1-\mathbf{v}_{1}^{2}} \; \mathbb{A}$, $\tilde{\mathbb{B}} \equiv \sqrt{1-\mathbf{v}_{2+}^{2}} \; \mathbb{B}$, and $\mathbb{I}_3$ is the $3 \times 3$ identity matrix.   
\par
In Eqs. (\ref{ort1}) and (\ref{ort2}), vectors $ \mathbf{\tilde F}_1$ and $ \mathbf{\tilde F}_{2}$ depend continuously on both history segments positions, velocities and accelerations.  
Near small-delay-angle circular orbits the separation in light-cone $r_{12} \equiv |\mathbf{x}_1-\mathbf{x}_{2+} |  \simeq r_b $ is large and $\mathbb{\tilde{A}}$ and $\mathbb{\tilde{B}}$ are $O(\frac{1}{r_{12}})$, such that for $r_{12}^2 \gg \frac{1}{m_1 m_2}$ the matrices on the left-hand-sides of (\ref{ort1}) and (\ref{ort2}) are non-singular quasi-diagonal matrices that can be inverted, yielding a Lipshitz-continuos non-autonomous ODE for the accelerations. 
\par
The dominant linear dependence on accelerations is obtained from the far-field component of (\ref{electric}) in the approximation of Eq. (\ref{far-rot}), which yields  $\tilde{\mathbb{A}}=\tilde{\mathbb{B}}=(1/r_{12})\tilde{\mathbb{Q}}$ with $\tilde{\mathbb{Q}}$ an $O(1)$ symmetric $3 \times 3$ matrix depending only on the normal along the light-cone. 
\par
Last, points $ \mathbf{x}_{2+}$ and $\mathbf{x}_1$ should evolve in the light-cone condition, and given that $\mathbf{a}_{2} \equiv d \mathbf{v}_{2+}/dt_{2+} $ and $\mathbf{a}_{1} \equiv d \mathbf{v}_{1}/dt_{1} $, a further transformation using $dt_{2+}/dt_1$ as given by (\ref{t2difft1}) is necessary to make both evolution parameters of (\ref{ort1}) and (\ref{ort2}) equal (a near-identity transformation). Equations (\ref{ort1}) and (\ref{ort2}) must then be used with a two-point boundary problem by choosing initial velocities such that orbits starting from  $\mathbf{O}_1$ and $\mathbf{O}^{+}_1$ hit $\mathbf{L}_2^-$ and $\mathbf{L}_2$. 
\par
  (ii) The two-point boundary problem uses up all the adjustable initial-positions and initial-velocities for ODE (\ref{ort1}) and (\ref{ort2}), and there is no freedom left to adjust that the \emph{end velocities} are continuous with history velocities at $\mathbf{O}_1^+$
and $\mathbf{L}^{-}_2$. The case of perfectly circular segments is  exceptional due to the existence of the circular solutions \cite{Schild}. From circular boundary data, the above integration simply continues the $C^\infty$ circular solution. Otherwise, from generic near-circular boundary data, the integration defines a near-circular orbit with velocity discontinuities at $\mathbf{O}^{+}_1$ and $\mathbf{L}^{-}_2$.  \; \; \; \; \qed
\par
As a bonus, the above construction shows that solutions with discontinuous velocities are \emph{expected}. For purely $C^2$ segments there are no Weierstrass-Erdmann conditions for shortest-length boundaries. Still for the shortest-length case, the above result can be generalized for boundary segments that are continuous and piecewise smooth, as follows. 
\par
For piecewise smooth boundary data, the ODE integration has to be stopped at every breaking point to satisfy the Weierstrass-Erdmann conditions (\ref{WE1}) that can be written as 
 \begin{eqnarray}
  \frac{m_{1}\Delta \mathbf{{v}}_{1}  }{ \sqrt{1 - \mathbf{v}_1^2}  }& =& \frac{\mathbb{G}_1 \Delta \mathbf{v}_{2+}}{r_{12}(1-(\mathbf{n} \cdot \mathbf{v}_{2+})^2)} +  \mathbb{U}_{2-} \Delta \mathbf{v}_{2-},   \label{Wrt1}\\
   \frac{m_{2}\Delta \mathbf{{v}}_{2+}  }{ \sqrt{1 - \mathbf{v}_{2+}^2}  }&=& \frac{\mathbb{ G}_2 \Delta \mathbf{v}_{1}}{r_{12}(1-(\mathbf{n} \cdot \mathbf{v}_{1})^2)}+ \mathbb{U}_{1+}\Delta \mathbf{v}_{1+} , \; 
\label{Wrt2}
\end{eqnarray} 
as obtained substituting $\mathbf{v}^r_{j\pm}= \mathbf{v}^l_{j\pm} + \Delta \mathbf{v}_{j\pm}$ into Eq. (\ref{WE1}). In Eqs. (\ref{Wrt1}) and (\ref{Wrt2}), $\mathbb{{G}}_1$ and $\mathbb{{G}}_2$ are $O(1)$ $3 \times 3$ matrices and we have explicitly introduced the extra factors in the denominators, which should be near-one for low-velocity orbits. Still in Eqs. (\ref{Wrt1}) and (\ref{Wrt2}), matrices $\mathbb{U}_{2-}$ and $\mathbb{U}_{1+}$ are bounded and depend continuously on the boundary segments positions, velocities and accelerations. Equations (\ref{Wrt1}) and (\ref{Wrt2}) can be solved for the velocity discontinuities along the unknown orbital segments, $\Delta \mathbf{v}_{1}$ and $\Delta \mathbf{v}_{2+}$, yielding
\begin{eqnarray}
 ( m_1 m_2 \mathbb{I}_3  - \lambda \mathbb{ G}_1 \mathbb{G}_2) \Delta \mathbf{v}_{1}& = &\mathbb{\tilde K}_{11} \Delta \mathbf{v}_{2-} +\mathbb{\tilde K}_{12} \Delta \mathbf{v}_{1+},   \label{danese1}  \\
  ( m_1 m_2 \mathbb{I}_3  - \lambda \mathbb{ G}_2 \mathbb{G}_1) \Delta \mathbf{v}_{2+}& = & \mathbb{\tilde  K}_{21} \Delta \mathbf{v}_{2-} +\mathbb{\tilde K}_{22} \Delta \mathbf{v}_{1+}  ,   \label{danese2}  \end{eqnarray} 
with
\begin{eqnarray}
\lambda &\equiv & \frac{\sqrt{1-\mathbf{v}_1^2} \sqrt{1-\mathbf{v}_{2+}^2} }{r^2_{12}(1-(\mathbf{n} \cdot \mathbf{v}_{1})^2)(1-(\mathbf{n} \cdot \mathbf{v}_{2+})^2)}.  \label{lambida}
\end{eqnarray}
\par
\textbf{Theorem II:} For continuous and piecewise $C ^2 $ boundary segments of shortest type, having a \emph{finite} number of velocity discontinuities and sufficiently close to circular segments of $\lambda \ll m_1 m_2 $ ($\lambda$ defined in Eq. (\ref{lambida})), the boundary value problem is well-posed in the $\widehat{C} ^2$ topology. 
\par
\textbf{Proof:} Whenever the ODE integration of Theorem I is halted because of a velocity discontinuity in a history segment, the $\Delta \mathbf{v}_{1+}$ and $\Delta \mathbf{v}_{2-}$ on the right-hand-side of  Eqs. (\ref{danese1}) and (\ref{danese2}) are small because boundary segments are sufficiently close to circular segments. Given that $\lambda \ll m_1 m_2 $, the matrices on the left-hand-sides of Eqs. (\ref{danese1}) and (\ref{danese2}) can be inverted, yielding small values for $\Delta \mathbf{v}_{1}$ and $\Delta \mathbf{v}_{2+}$. Eventually, the two-point-boundary-value problem yields an orbit still close to the circular segment. \; \; \; \;  \qed
 \par
 The above constructed continuous trajectories have as many velocity discontinuities as the combined past/future history segments have.
 Theorem II generalizes to boundary segments near constant-velocity-straight-line-segments at large separations and small velocities. For both circular and straight-line boundary segments, matrices $\mathbb{\widehat K}_{ij}$ on the right-hand-side of Eqs. (\ref{danese1}) and (\ref{danese2}) fall as $1/{r_{12}}$ (not indicated), such that universal perturbations of distant charges decay with distance, as mentioned at the end of Section \ref{Section IV}.
  \par
 Notice that the quantity $\lambda$ defined in Eq. (\ref{lambida}) appeared earlier in Eq. (\ref{stepping-stone2}) of Section \ref{Section VII} in a completely different limit.
 For the periodic orbit of Section \ref{Section VII}, the matrices on the left-hand-side of Eqs. (\ref{danese1}) and (\ref{danese2}) would be near-singular because the stepping-stone condition is $\lambda \approx m_1 m_2 $, but then the velocity jumps on the right-hand-sides of (\ref{danese1}) and (\ref{danese2}) are not arbitrary because in Section \ref{Section VII} we are dealing with a  periodic orbit.  
 \par
 The problem of boundary segments with longer time-spans requires inversion of larger matrices. Figure \ref{Squema} illustrates a longer boundary-value problem $(\Delta t_j \simeq 3r_b)$. Notice in FIG. \ref{Squema} that sewing chains starting from points either on segment $(\mathbf{O}_1, \mathbf{b}_2) $ or on segment $(\mathbf{f}_1, \mathbf{L}^{-}_2)$ have \emph{two} vertices along trajectory $1$, while sewing chains starting from points on the central segment $(\mathbf{b}_2,\mathbf{f}_1)$ have \emph{just one} vertex along trajectory $1$. The situation of trajectory $2$ is analogous. The Wheeler-Feynman advance/delay equations  for accelerations $\mathbf{a}_{11}, \mathbf{a}_{12}, \mathbf{a}_{21}, \mathbf{a}_{22}$ illustrated in FIG. \ref{Squema}, are obtained analogously to the case of Theorem I, yielding 
  \begin{equation}
  \left( \begin{smallmatrix} m_1 \mathbb{I}_3 & 0 & \frac{\mathbb{G}_a}{r_b} & 0 \\  0 & m_1 \mathbb{I}_3& \frac{\mathbb{G}_b}{r_b}& \frac{\mathbb{G}_c}{r_b} \\ \frac{\mathbb{G}_a}{r_b} & \frac{\mathbb{G}_b}{r_b} &m_2 \mathbb{I}_3 & 0\\ 0 & \frac{\mathbb{G}_c}{r_b} & 0 & m_2 \mathbb{I}_3  \end{smallmatrix} \right)
  \left(\begin{smallmatrix} \mathbf{a}_{11} \\ \; \\ \mathbf{a}_{12} \\ \;  \\  \mathbf{a}_{21}\\ \;  \\ \mathbf{a}_{22}   \end{smallmatrix} \right)=\mathbb{F}(t_1,\mathbf{X},\mathbf{V}, boundary \; segments) , \label{matrizaa} 
\end{equation} 
where $\mathbf{X}$ and $\mathbf{V}$ indicate positions and velocities of the four running vertices of the sewing chain illustrated by solid violet lines in FIG. \ref{Squema}. 
\par
\textbf{Theorem III:} For near-circular $C^2$ boundary segments with $ 2r_b < \Delta t_j < 4r_b$, the unique solution depends continuously on the boundary data (with the $\widehat{C}^2$ topology) and has two velocity discontinuities \emph{inside} each trajectory of the solution segment,  points $(\mathbf{b}_1, \mathbf{f}_2)$ (green) and $(\mathbf{b}_2, \mathbf{f_1})$ (blue) of FIG. \ref{Squema}. 
\par 
\textbf{Proof:} In Eq. (\ref{matrizaa}), $\mathbb{G}_a, \mathbb{G}_b$ and $\mathbb{G}_c $ are $O(1)$ symmetric $3 \times 3 $ matrices, just like in Theorem I. Explained as an initial-value problem illustrated by violet arrows in FIG. \ref{Squema}, after the non-singular near-diagonal $12 \times 12 $ symmetric matrix on the left-hand-side of Eq. (\ref{matrizaa}) is inverted (for small $\lambda$), integration of ODE (\ref{matrizaa}) should start from $\mathbf{x}_{11}(0)=\mathbf{x}_{{O}_1}$, $\mathbf{x}_{21}(0)=\mathbf{x}_{O^+_1}$, $\mathbf{x}_{12}(0)=\mathbf{x}_{{f}_1}$, $\mathbf{x}_{22}(0)=\mathbf{x}_{f_2}$. 
\par
The initial near-circular velocities and the initial positions $(\mathbf{f}_1,\mathbf{f}_2)$ are not known unless for circular boundary segments. Otherwise, for near-circular segments these must be chosen such that at the end-point $(\mathbf{x}_{12},\mathbf{x}_{22})=(\mathbf{L}^{-}_2,\mathbf{L}_2)$ and the running light-cone-ray $(\mathbf{x}_{11}, \mathbf{x}_{21})$ is ray $(\mathbf{b}_2,\mathbf{b}_1)$ of the backwards sewing chain of $\mathbf{L}^{-}_2$. 
\par
The remaining central segments $(\mathbf{b}_1, \mathbf{f}_2)$ (green) and $(\mathbf{b}_2, \mathbf{f_1})$ (blue) of FIG. \ref{Squema} are done in the manner of Theorem I, generating velocity discontinuities at $\mathbf{f}_1$, $\mathbf{f}_2$, $\mathbf{b}_1$ and $\mathbf{b}_2$, which must satisfy Weierstrass-Erdmann conditions, one over each orbital corner of each principal sewing chain.  Counting the end-point velocity discontinuities at $\mathbf{O}^+_1$ and $\mathbf{L}^{-}_2$, the generic case has \emph{six} velocity discontinuities even for $C^2$ boundaries.  \;\;\; \qed
 \par
 The above theorems suggest that the variational problem makes sense at least in the neighbourhood of circular orbits \cite{Schild}. 
 Notice that Eq. (\ref{matrizaa}) is a \emph{non-autonomous} ODE because the advanced and retarded arguments depend explicitly on the running time $t_1$ via the light-cone condition (\ref{light-cone}). The well-posedness of the general boundary-value problem is an open problem.
 \par
 Last, the former theorems predicted a critical distance $r_{12}^2 \simeq \frac{1}{m_1 m_2}$ below which the matrices can no longer be inverted and the equations of motion are differential-algebraic \cite{Petzold}. It is interesting to notice that the critical magnitude is of the order of the nuclear magnitude.

\section{Generalized Wheeler-Feynman electrodynamics}
\label{Section IV}

 Notice that $\mathbf{A}_j$ is defined by Eq. (\ref{preWE1}) only \emph{on} points of the finite segment of trajectory $i$ illustrated in FIG. \ref{Squema}. To extend $\mathbf{A}_j$ to a field $\mathbf{A}_j(t,\mathbf{x})$ we need $t_{j-} <  t < t_{j+}$ for some $t_{j-}$ and $t_{j+}$ belonging to the finite segment of trajectory $j$ illustrated in FIG. \ref{Squema}.  For example, for $j=2$ the light-cone distances defined by Eq. (\ref{fieldrj}) evaluate to $r_{2-}(t,\mathbf{x})=t-t_{2-}$ and $r_{2+}(t,\mathbf{x})=t_{2+}-t$, which implies that $r_{2-}+r_{2+}=t_{2+}-t_{2-} < t_{{L_{2}}}-t_{O_1^-} $ (see FIG. \ref{Squema}). An analogous consideration shows that one can extend $\mathbf{A}_1$ to a field only when $\mathbf{x}$ is within a finite distance of the segment of trajectory $1$ illustrated in FIG. \ref{Squema}. The region of  common extension is the intersection $\mathcal{B}$ of the former two bounded regions of space-time. 
 \par
 Inside $\mathcal{B}$, the electromagnetic fields of both particles are naturally extended with the Li\'{e}nard-Wiechert formulas (\ref{electric}) and (\ref{magnetic}), just like in Wheeler-Feynman electrodynamics \cite{Whe-Fey}. Notice that the extended fields are undefined for points in the light-cone relation with corner points because then the past/future velocities and accelerations of the other charge are not defined.
\par
The above considerations suggest a generalized Wheeler-Feynman electrodynamics restricted to the bounded region of space-time $\mathcal{B}$ \cite{Whe-Fey}, using the
\emph{finite} segments of trajectories provided by the critical points of the variational principle of Section \ref{Section II} \cite{JMP2009, Minimizers}. In the same way as in Wheeler-Feynman electrodynamics \cite{Whe-Fey}, formulas (\ref{electric}) and (\ref{magnetic}), which are borrowed from the equations of motion (\ref{Lorentz}) \emph{along} trajectories, are used to extend the fields inside $\mathcal{B}$, modulo some sets of zero volume in light-cone with the breaking points. 
\par
If a trajectory has a discontinuity at point $P_D \equiv (t_D, D)$, its extended fields at time $t $ are undefined on the critical sphere $S_D$ of radius $r_D \equiv |t-t_D|$ (the set of points either in the past or in the future light-cone of $P_D$). The experimentally verified \emph{integral} laws of classical electrodynamics are recovered in the following way. Gauss's law involving the surface integral of the electric field at time $t$ holds if/when (i) the Gaussian surface $G$ is inside $\mathcal{B}$ and (ii) the critical spheres emanating from each discontinuity point inside $G$ intersect $G$ on sets of zero measure. Any surface $S_{accept.} \in \mathcal{B} $ intersecting $S_D$ along a set of zero measure is acceptable, e.g., the surface of a cube.
\par
The proof of Gauss's law is exactly that of Wheeler and Feynman \cite{Whe-Fey} using the following functional-theoretic density argument \cite{Brezis}. A piecewise $C^2$ continuous trajectory $T_{\infty}$ with a \emph{finite} number of velocity discontinuities can be recovered as the limit of a sequence $T_n$ of $C^2$ trajectories whose extended fields satisfy Gauss's integral law for $S_{accept.} \in \mathcal{B}$. The surface integral survives the limit if the former conditions (i) and (ii) hold. As in Wheeler-Feynman electrodynamics \cite{Whe-Fey}, the surface integral of the electric field over $S_{accept.}$ is equal to the charge inside $S_{accept.}$, while the surface integral of the magnetic field over $S_{accept.}$ vanishes (as usual in electrodynamics).
\qed
\par
Last, for the special case when variational trajectories plus boundary segments form $C^2$ segments, also the \emph{differential} form of Maxwell's equations holds inside $\mathcal{B}$, as proved in the manner of Wheeler and Feynman \cite{Whe-Fey}.
\par
Extension of electromagnetic fields to almost everywhere in $\mathbb{R} \times \mathbb{R}^3$ (time$\times$space) requires extremal trajectories defined in $t \in [-\infty, \infty]$, henceforth \emph{globally defined}. As long as trajectories have a \emph{finite number} of corners per finite segment, formulas (\ref{electric}) and (\ref{magnetic}) define extended fields almost everywhere but for a finite number of surfaces in $\mathcal{B}$, which are sets of zero measure (volume).
\par
The same generalizations carry over for Ampere's integral law and Faraday's induction law in integral form \cite{Jackson}, as obtained by restricting the proofs of Wheeler and Feynman \cite{Whe-Fey} to curves and surfaces of $\mathcal{B}$ having finitely measured intersections with the relevant critical spheres $S_D$.  Results following from laws in differential form do not carry over from Maxwell's electrodynamics to variational electrodynamics. For example, Poynting's theorem is valid only in regions where extended fields are $C^2$ \cite{Minimizers}.    
 \par
 Extended fields of trajectories with an \emph{infinite} number of corners per finite segment would require a Lebesgue integral to define the action, and are not studied here. Generalizations of the integral laws of classical electrodynamics using Sobolev's trace theorems \cite{Brezis}, and a variational principle using Lebesgue-integrable action functionals, are open problems.
\par
\begin{figure}[h!]
   \centering
  \includegraphics[scale=0.5]{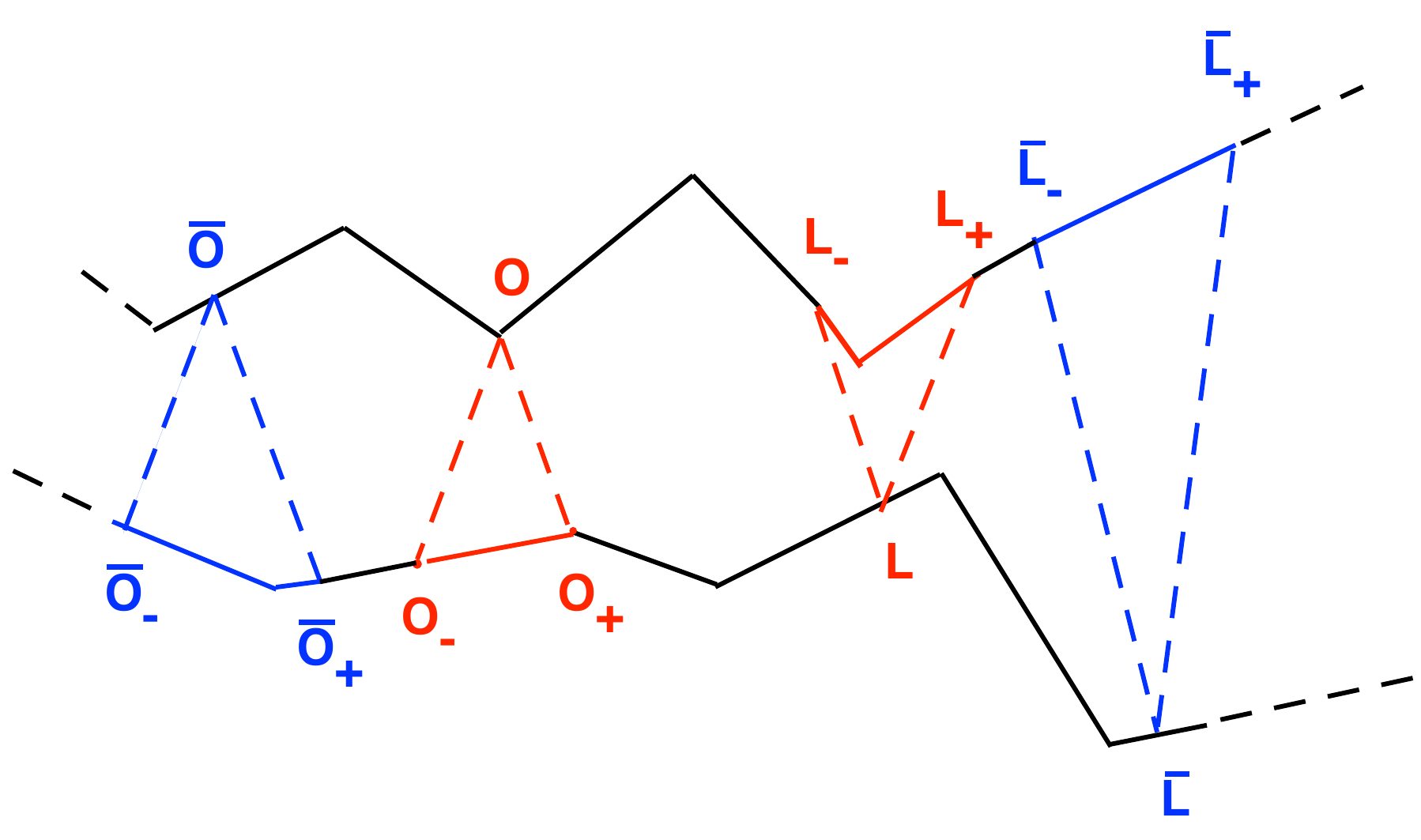} 
\caption{Illustration of a continuous and \emph{piecewise} smooth invariant manifold, along with two different sets of boundary segments (blue triangles and red triangles). }
  \label{fig7}
\end{figure}

 An invariant manifold $\mathcal{M}$ of the variational two-body problem is a pair of continuous and \emph{piecewise} smooth trajectories such that for any pair of boundary segments in $\mathcal{M}$, the extremum of the corresponding boundary-value problem is a segment of $\mathcal{M}$, as illustrated in FIG. \ref{fig7}. Unlike the case of an ODE, trajectories of $\mathcal{M}$ can have corners and one can \emph{not} continue trajectories with a time integration (neither forward nor backwards). The infinite-dimensional problem at hand is to find a whole function, just like solving a partial differential equation (PDE) with boundary conditions.  

\par
For a many-body system, physically interesting are \emph{globally bounded} invariant manifolds. Given that all trajectories are spatially bounded, one can show that all advanced and retarded normals coincide at a far distance, i.e., $ \mathbf{n}_{j+}\rightarrow \mathbf{n}$ and $ \mathbf{n}_{j-} \rightarrow \mathbf{n}$, $ \forall j$ \cite{Minimizers}. The extended far-electric field thus becomes
\begin{equation}
\mathbf{E}(t,\mathbf{x})=\frac{1}{2}\sum_j (\mathbf{E}_{j+}(t,\mathbf{x})+ \mathbf{E}_{j-}(t,\mathbf{x}) ), \label{semisum}
\end{equation}%
while Eq. (\ref{magnetic}) with $ \mathbf{n}_{j+} = \mathbf{n}_{j-} \equiv \mathbf{n}$ yields the extended far-magnetic field (i.e., for $|\mathbf{x}| \rightarrow \infty$)
\begin{equation}
\mathbf{B}(t,\mathbf{x})=\frac{1}{2} \mathbf{n} \times \sum_j (\mathbf{E}_{j-}(t,\mathbf{x})- \mathbf{E}_{j+}(t,\mathbf{x}) ) \label {semidiff}.
\end{equation}%
\par
The many-body system (e.g. a multi-electron atom) is \emph{isolated} from its far surroundings when an extra distant charge can travel undisturbed at a far distance with an \emph{arbitrary} velocity $\mathbf{v}_i$. The necessary isolating condition is that the electric and magnetic extended far fields on the right-hand-side of force law (\ref{Lorentz}) should vanish almost everywhere at a far-distance, i.e., the \emph{semi-sum} (\ref{semisum}) and \emph{semi-difference} (\ref{semidiff}) should vanish asymptotically. The vector product with $\mathbf{n}$ in Eq. (\ref{semidiff}) amounts to no extra freedom because far-fields are transversal. The former isolating condition is essentially Wheeler and Feynman's \emph{absorber condition} \cite{Whe-Fey} generalized to continuous and piecewise smooth globally bounded extrema.  
 \par
Last, about the $n$-charge problem: The corresponding action functional (\ref{aFokker}) has $n$ terms of type $T_i$ and $n(n-1)$ interactions between pairs, $V^{\pm}_{ij}$. Just like in Wheeler-Feynman electrodynamics \cite{Whe-Fey}, charges contribute linearly to the extended fields with a term that falls at the most with $(1/r)$. The contribution to the Weierstrass-Erdmann condition (\ref{WE1}) also falls at the most with $(1/r)$, wherever defined in space-time. It is therefore a good approximation to disregard universal perturbations of distant charges.
\section{ Circular Orbits}
\label{Section V}

The electromagnetic two-body problem has globally defined circular orbit solutions \cite{Schild}. The stability of circular orbits is studied in \cite{astar2B, Hans, pre_hydrogen} and the quantization of circular orbits is discussed in \cite{Hans_relativistic, HansWKB}. Circular orbits with large radii are discussed below using the notation of Ref. \cite{pre_hydrogen}.
\par
 The constant angular velocity and distance in light-cone are denoted by $\Omega $
and $r_{b}$, respectively,  and the angle $\theta \equiv \Omega r_{b}$ that particles turn in the light-cone time is henceforth called the \textit{delay angle}. 
The family of subluminal circular orbits \cite{Schild} is parametrized by $\theta $, and for
quantum Bohr orbits it turns out that $\theta \lesssim 10^{-2}$ \cite{Bohr,Bethe}. Along orbits of a small delay angle the Kepler formulas yield the leading
order angular velocity and distance in light-cone, respectively
\begin{eqnarray}
\Omega &=&\mu \theta ^{3}+O(\theta ^{5}) ,  \label{Omega} \\
r_{b}&=&\frac{\theta}{\Omega}=\frac{1}{\mu \theta ^{2}},   \label{RB}
\end{eqnarray}%
where reduced mass and total mass are defined by $\mu \equiv m_{1}m_{2}/M$ and $M\equiv m_{1}+m_{2}$. 
It is important to keep these limiting dependencies in mind, and for hydrogen $\mu  \simeq (1836/1837)\simeq 1$.
Adopting the same notation of Ref. \cite{pre_hydrogen}, we express each particle's circular orbit's radius by numbers $0 \leq b_i < 1$  as
\begin{equation}
r_{i}  \equiv  b_{i}r_{b},  \label{defradius} \\
\end{equation}%
which define scalar velocities 
\begin{equation}
v_{i} =\Omega r_{i}=\theta b_{i}, \label{defvelocity}
\end{equation}
for $i=1,2$.

\begin{figure}[here]
   \centering
    \includegraphics[scale=0.45]{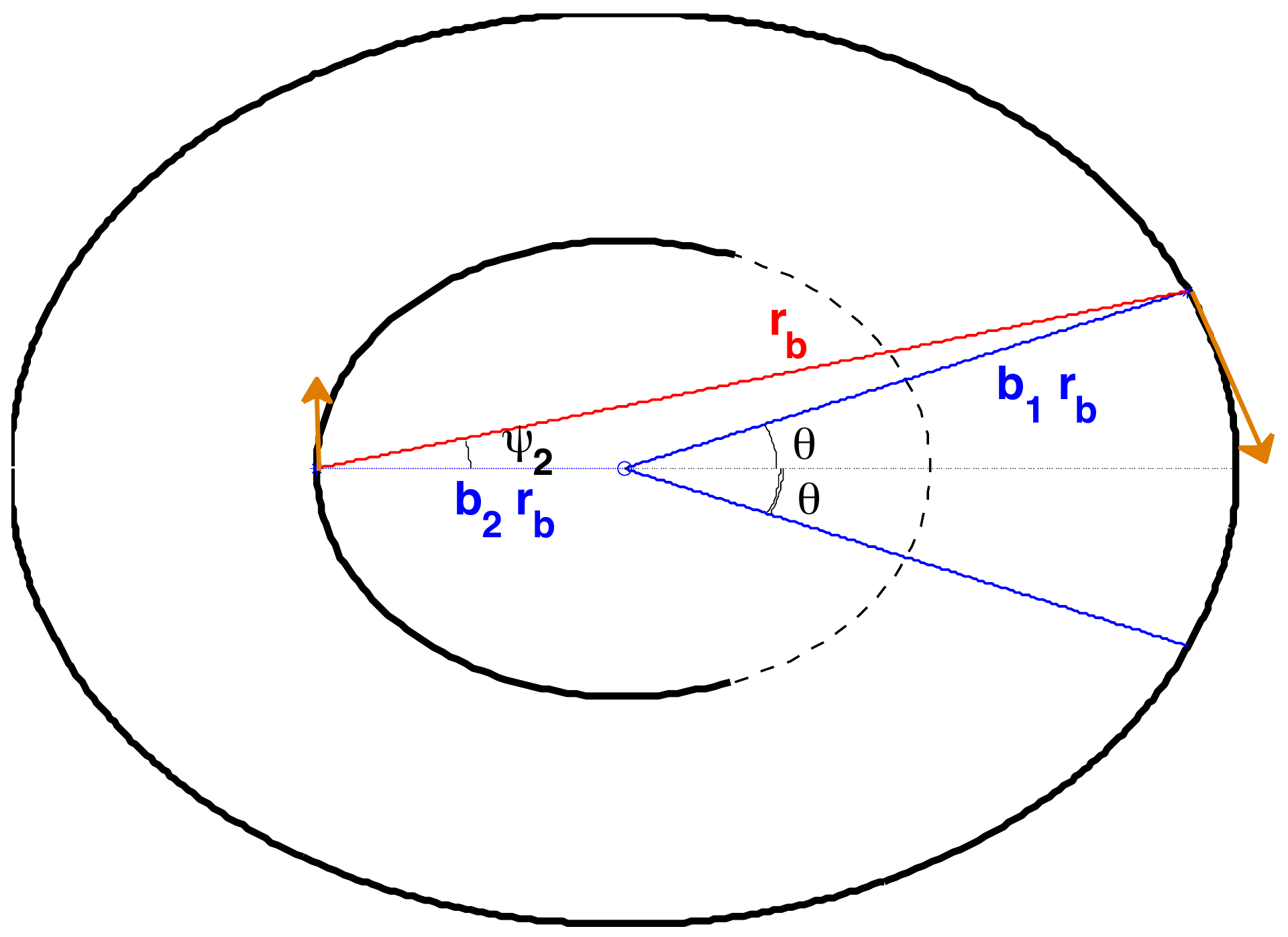} 
  \caption{Schild circular orbit with particles in diametrical opposition at the same time of the inertial frame ($x$-axis, dotted straight line is $t=0$). Indicated are the advanced/retarded positions of particle 1 along the outer circle, and the (equal) angle(s) $\theta $ travelled during the past/future light-cone time intervals.  Light-cone distance $r_b$ in the past light-cone of particle 2 is the red line making an angle $\psi_2$ with the horizontal line. Arbitrary units. 
  }
   \label{fig1}
\end{figure}
 The speed of light limit $c \equiv 1$ imposes that $\max(\theta b_1,\theta b_2) \leq 1 $. In the limit of a small mass-ratio, $(\mu/M) \rightarrow  0$, one has $\theta \in (0,1]$, the upper limit $\theta=1$ corresponding to the particle of smaller mass ($m_1$) traveling at the speed of light. As illustrated in FIG. \ref{fig1} and in Ref. \cite{Schild}, the circular radii and the distance
in light-cone form a triangle of largest side $r_{b}$, yielding a trigonometric constraint
\begin{equation}
b_{1}^{2}+b_{2}^{2}+2b_{1}b_{2}\cos (\theta )=1,
\label{circular_cone}
\end{equation}
equivalent to Eq. (3.1) of Ref. \cite{Schild}.
Using the ratio of equations (3.2) and (3.3) of Ref. \cite{Schild} to eliminate $\Omega$, together with constraint (\ref{circular_cone}), 
we find  the solution 
 $b_{1}= (1+\frac{\mu \theta^2 }{2M}) \frac{m_2}{M} +O(\theta ^{4})$ and  $
b_{2}= (1+\frac{\mu \theta^2 }{2M})\frac{m_1}{M}+O(\theta ^{4})$. 
\par
Last, the angular momentum in units of electronic charge squared over speed of light $(e^2/c)$,
\begin{equation}
l_{z}=\frac{1+b_1 b_2 \theta^2 \cos(\theta)}{\theta +b_1 b_2 \theta^2 \sin(\theta)} \simeq \frac{1}{\theta}, \label{angular-momentum}
\end{equation}%
is an important quantity of the circular orbit to keep in mind \cite{Schild, Hans, Hans_relativistic}. Atomic orbits
have $l_{z}$ of the
order of one over the fine-structure constant, $ \hbar c/ e^2=137.036$, a fundamental magnitude of atomic physics. Delay angles and angular momenta of Bohr orbits are respectively $ \theta \simeq   ( e^2/\hbar c) / q$ and $l_z \simeq  (\hbar c/e^2) q $, for each nonzero integer $q$ \cite{Bohr, Sommerfeld}.
\par

\section{Linearization about circular orbits}
\label{Section VI}

In order to linearize the equations of motion about circular orbits, we re-write the Lorentz-force Eq. (\ref{Lorentz}) as 
\begin{equation}
\frac{m_i \mathbf{a}_{i
}}{\sqrt{1-\mathbf{v}_{i}^{2}}}= e_i [ \mathbf{E}_j-(\mathbf{v}_i \cdot \mathbf{E}_j)\mathbf{v}_i+ \mathbf{v}_{i
} \times \mathbf{B}_j] \label{Hans}, 
\end{equation}
by evaluating the derivative on the left-hand side of Eq. (\ref{Lorentz}) and subtracting the scalar product with $\mathbf{v}_i$, where again $i=1,2$ and $j  \equiv 3-i $. The magnetic term (last term on the right-hand side of (\ref{Hans})) is a transversal force proportional to the electric field by (\ref{magnetic}), and further proportional to the velocity modulus $|\mathbf{v}_i|$, which is small for small $\theta$ (see Eq. (\ref{defvelocity})). 
\par
 For small delay angles, the  electric force, first term on the right-hand-side of Eq. (\ref{Hans}), is responsible for the significant contributions to the linearized equations along a circular orbit. The electric field (\ref{electric}) of charge $j \equiv3-i$ decomposes in two terms: (i) the near-electric field proportional to $1/r_{j\pm}^2$, a magnitude of $\mu^2 \theta^4$ by use of (\ref{RB}), times a unit vector (almost) along the particle separation at the same time, $\mathbf{n}(t)$; and (ii) the far-electric field proportional to $1/r_{j\pm}$, a magnitude of $\mu \theta^2$ by use of (\ref{RB}), times a vector along the circular orbit.
  \par
 It is important to ponder upon the magnitude of each electric contribution: the far-electric field is almost transversal to the particle separation at the same time, $\mathbf{n}(t)$, such that projection along $\mathbf{n}(t)$ involves the (small) factor of $\cos(-\theta + \pi/2) \simeq \theta$.  Separating the contributions of near-fields and far-fields, the right-hand-side of  (\ref{Hans}) has the combined magnitude
\begin{eqnarray}
F_i \propto O(\theta^4) + |\mathbf{a}_j| O(\theta^3). \label{magnitudes}
\end{eqnarray}
In Eq. (\ref{magnitudes}), the near-field contribution is $O(\theta^4)$ while the far-field contribution is $ |\mathbf{a}_j | O(\theta^3)$, which is smaller along circular orbits since $ |\mathbf{a}_j| \propto 1/{r}_b ^2 = O(\theta^4)$. The force along circular orbits is approximately described by the near field only \emph{because} the far-field (\ref{electric}) is further proportional to the $O(\theta ^4)$ acceleration of circular orbits. 
Upon linearization about the circular orbit, this dominance changes: the linearized equations accept solutions of arbitrarily large accelerations, and the most important contribution to the linearized version of Eq. (\ref{Hans}) is precisely the contribution of the \emph{far-field term}.
\par
Next we derive the linearized equations along the orbital plane using the notation of Ref. \cite{pre_hydrogen}.  We introduce complex gyroscopic coordinates where the circular orbit is a fixed point of the equations of motion, i.e.,
\begin{equation}
x_{j}+iy_{j}\equiv r_{b}\exp (-i\Omega t)[b_{j}+l_j+iu_j],
\label{coordinates} 
\end{equation}%
where $(l_j , u_j)$ are respectively the longitudinal and transversal gyroscopic coordinates. The circular orbit \cite{Schild} is the fixed point $(l_j,u_j)=(0,0)$ for $j=1,2$. Again, while small-delay-angle circular orbits have $O(\theta^4)$ accelerations, the linearized acceleration corrections, $\mathbf{\delta a}_j $, can be arbitrarily large \cite{pre_hydrogen}.  The dominant linear correction for the accelerations (the stiff limit of Ref. \;\cite{pre_hydrogen}) is obtained using (\ref{Hans}) with only the first term on the right-hand side and far-field $\mathbf{E}_{j\pm} $ approximated by 
\begin{equation}
\mathbf{E}_{j\pm} \simeq \frac{1}{r_b} \mathbf{n}_{j\pm} \times (\mathbf{n}_{ j\pm} \times \mathbf{\delta a}_{j\pm} )=\frac{1}{r_b}(\mathbf{n}_{j\pm} \cdot \mathbf{\delta a}_{j\pm})\mathbf{n}_{j\pm}-\frac{1}{r_b}\mathbf{\delta a}_{j\pm} \label{far-rot},
\end{equation}
where $\pm$ indicate evaluation at the unperturbed deviating times $t \pm (\theta /\Omega) $ (because we want the linear term only).
The gyroscopic representation of the rotating normals in light-cone are complex numbers of unit modulus, i.e., $\exp (-i\Omega t) \exp({i \psi_j})$, where 
\begin{eqnarray}
\exp(i \psi_1)  \equiv b_1 + b_2 \exp{(i \theta)}, \notag \\
\exp(i \psi_2)  \equiv b_2 + b_1 \exp{(i \theta)}. \label{cosntr}
\end{eqnarray}
Equation (\ref{circular_cone}) can be used to show that the modulus of each complex number on the right-hand-side of (\ref{cosntr}) is \emph{unitary}.  Angles $\psi_1 $ and $ \psi_2$ 
further satisfy $\psi_1 + \psi_2 =\theta$, and for small $\theta$ the $b_i$ given below Eq. (\ref{circular_cone}) yield $\psi_1 \simeq \frac{m_1}{M} \theta$ and $\psi_2 \simeq \frac{m_2}{M} \theta$. In FIG. \ref{fig1}, $\psi_2$ is the angle between the (rotating) red line and the $x$-axis (dashed line).
\par
The linearized planar equations of motion are obtained substituting (\ref{coordinates}) into (\ref{Hans}) with only its first right-hand-side term given by the semi-sum of retarded and advanced fields (\ref{far-rot}). The real and complex parts of the linearized equations of motion keeping only the largest derivatives of the gyroscopic coordinates yields 
\begin{eqnarray}
 \negthickspace \negthickspace  \negthickspace \negthickspace \negthickspace m_1 r_b \ddot{l}_1 &=  &\negthickspace - C_{11}\frac{(\ddot{l}_{2+} +\ddot{l}_{2-})}{2}- {S_{11}}\frac{(\ddot{u}_{2+}-\ddot{u}_{2-})}{2}, \notag \\ 
 \negthickspace \negthickspace  \negthickspace \negthickspace \negthickspace m_2 r_b \ddot{l}_2 &=  &\negthickspace -C_{21}\frac{(\ddot{l}_{1+} +\ddot{l}_{1-})}{2}- S_{21}\frac{(\ddot{u}_{1+}-\ddot{u}_{1-})}{2}, \notag \\
\negthickspace \negthickspace  \negthickspace \negthickspace \negthickspace m_1 r_b \ddot{u}_1 &= &\negthickspace - C_{31} \frac{(\ddot{u}_{2+} +\ddot{u}_{2-})}
  {2} -S_{31}\frac{(\ddot{l}_{2+}-\ddot{l}_{2-})}{2}, \notag \\ 
 \negthickspace \negthickspace  \negthickspace \negthickspace \negthickspace m_2 r_b \ddot{u}_2 &=  &\negthickspace - C_{41}\frac {(\ddot{u}_{1+} +\ddot{u}_{1-})}{2}-S_{41}\frac {(\ddot{l}_{1+}-\ddot{l}_{1-})}{2}, \label{4eqs}
 \end{eqnarray}
where $C_{j1}$ and $S_{j1}$ are $4 \times 1$ matrices defined by
\begin{equation}
 C \equiv  - \left( \begin{smallmatrix}  \cos{\theta}-\cos\psi_2 \cos{\theta} \\  \cos{\theta}-\cos\psi_1 \cos{\theta} \\ \cos{\theta}+\sin{\psi_2}\sin{\theta} \\ \cos{\theta}+\sin{\psi_1}\sin{\theta} \end{smallmatrix} \right)  \equiv  (\cosh{\lambda})^{-1} D, \label{DD}
\end{equation} 
and 
\begin{equation}
S \equiv - \left( \begin{smallmatrix}   \sin{\theta}-\sin{\psi_2}\cos{\theta}  \\   \sin{\theta}-\sin{\psi_1}\cos{\theta} \\   (1-\cos\psi_2) \sin{\theta}   \\ (1-\cos\psi_1) \sin{\theta} \end{smallmatrix} \right) \equiv (\sinh{\lambda})^{-1} E. \label{DE}
\end{equation}
In Eqs. (\ref{DD}) and (\ref{DE}), $4 \times 1 $ matrices $D$ and $E$ are defined to be used below. 
\par
Equation (\ref{4eqs}) is a \emph{linear} NDDE  with exponential solutions
\begin{eqnarray}
l_j & =& L_j \exp(\lambda \Omega t /\theta), \notag \\
u_j & =& U_j \exp(\lambda \Omega t /\theta) \label{exponent},  
\end{eqnarray}
where $j=1,2$ and $(L_1,L_2,U_1,U_2 )$ 
is a non-trivial solution of 
\begin{equation}
  \left( \begin{smallmatrix} m_1 r_b & D_{11}& 0 &E_{11} \\  D_{12} & m_2 r_b& E_{12}& 0  \\ 0 & E_{13} &m_1 r_b &D _{13}\\ E_{14} & 0 & D_{14} & m_2 r_b  \end{smallmatrix} \right)
  \left(\begin{smallmatrix} L_1 \\  L_2 \\  U_1 \\ U_2   \end{smallmatrix} \right)=0. \label{matriz} 
\end{equation} 
\par
  A nontrivial solution of Eq. (\ref{matriz}) requires the vanishing of its $4 \times 4 $ determinant,
\begin{equation}
F_{xy}=1-\frac{\mu \theta^4}{M} \cosh^2(\lambda)=0, \label{detxy0}
\end{equation}
where powers of $\theta^2 \cosh{\lambda}$ with coefficients of $O(\theta^4)$ were discarded, henceforth and in Ref. \cite{pre_hydrogen} called the stiff-limit. The roots of Eq. (\ref{detxy0}) exist in symplectic sets of four, i.e., $(\lambda, -\lambda, \lambda^{*},-\lambda^{*})$. For atomic hydrogen, $\theta ^{-1}\sim 137.036$ and
$(\mu /M) \simeq (1/1836)$, such that Eq. (\ref{detxy0}) requires $\lambda $  to have a positive real part
$| \Re(\lambda )|\equiv  \sigma=  \ln (\sqrt{\frac{4M}{%
\mu \theta ^{4}}}) \simeq 14.29$. The
imaginary part of $\lambda $ is any integer multiple of $\pi i $. The general solution of Eq. (\ref{detxy0}) modulo the symplectic symmetry is
\begin{equation}
\lambda =\sigma +\pi qi,  \label{unperastar}
\end{equation}%
where $i\equiv \sqrt{-1}$ and $q \in \mathbb{Z}$. 
\par
Solution (\ref{unperastar}) was called a ping-pong mode in Ref. \cite{pre_hydrogen} because its phase advances by $\pi q$ in one light-cone time $r_b$, a phase speed of $\pi q \Omega/\theta =\mu \theta^2 \pi q= \pi q / r_b=(\Omega/ \theta) \Im({\lambda})$. The only $O(1)$ off-diagonal terms of matrix (\ref{matriz}) are $D_{13}$ and $D_{14}$, others being $O(\theta)$ or higher order. The nontrivial eigenvector solution is approximately 
\begin{eqnarray}
(L_1,L_2,U_1,U_2) \propto (\frac{\mu \theta}{M},\frac{\mu \theta}{M},1,\sqrt{\frac{\mu}{M}}) \label{eigenvector}.
\end{eqnarray}
\par
For $m_2 \gg m_1$, normal-mode solution (\ref{exponent}) oscillates (almost) along the circular orbit because Eq. (\ref{eigenvector}) yields $U_i \gg L_i$, thus defining a quasi-\emph{transversal} mode. The largest longitudinal component, $L_i$, is attained for positronium at moderate $\theta$. Solution (\ref{unperastar}) defines a nonzero real part for $\lambda$ that causes amplitudes to blow up at either $t \to \pm \infty $, implying that besides Schild orbits\cite{Schild}, no other almost-circular orbit can be simultaneously $C ^2$ \emph{and} globally bounded.  
\par
The inclusion of $O(1/\lambda)$ and $O(1/\lambda^2)$ linear terms to the planar motion is outlined in Ref. \cite{pre_hydrogen}.
The linearized motion perpendicular to the orbital plane is studied analogously.
As explained in \cite{pre_hydrogen}, the $z$-oscillations are transversal modes that decouple from the planar transversal oscillations (\ref{4eqs}) at linear order. The determinant of the linearized $2 \times 2$ system in the limit $|\lambda| \rightarrow \infty$ is
again  (\ref{detxy0}), an \emph{asymptotic} degeneracy.
The degeneracy is raised by $O(1/\lambda^2)$ corrections to (\ref{matriz}) and (\ref{detxy0}) introduced by the linear terms with lower derivatives \cite{pre_hydrogen}. 
\par
The determinant for planar modes is calculated in Ref. \cite{pre_hydrogen} up to $O(\frac{1}{\lambda^4})$ terms 
(see Eq. (41) of Ref. \cite{pre_hydrogen} with $\Gamma =-1/2$ ), i.e.,
\begin{eqnarray}
F_{xy}&=&1-\frac{\mu \theta^4}{M}(1+\frac{7}{\lambda^2}+\frac{5}{\lambda^4}) \cosh^2(\lambda)\nonumber \\&&+\frac{\mu \theta^4}{M}(\frac{1}{\lambda}+\frac{5}{\lambda^3})\sinh(2\lambda)=0, \label{Dexy}
\end{eqnarray}
where we have disregarded $O(\frac{\theta^2}{\lambda^2})$ terms not proportional to the large hyperbolic functions. The determinant for perpendicular oscillations and up to $O(\frac{1}{\lambda^4})$ terms is Eq. (B17) of appendix B in Ref. \cite{pre_hydrogen} with $\Gamma =-1/2$, i.e., 
\begin{eqnarray}
F_{z}&=&1-\frac{\mu \theta^4}{M}(1-\frac{1}{\lambda^2}+\frac{1}{\lambda^4}) \cosh^2(\lambda)\nonumber \\&&+\frac{\mu \theta^4}{M}(\frac{1}{\lambda}-\frac{1}{\lambda^3})\sinh(2\lambda) =0,\label{Dez}
\end{eqnarray}
 again disregarding $O(\theta^2)$ terms that are not multiplied by the large hyperbolic functions. Notice the symplectic symmetry that roots of Eqs. (\ref{Dexy}) and (\ref{Dez}) are still in sets of four, $(\lambda, -\lambda, \lambda^{*},-\lambda^{*})$. The first correction at $O(\frac{1}{\lambda})$ is the same for the roots of both of Eqs. (\ref{Dexy}) and (\ref{Dez}). The corrections at $O(\frac{1}{\lambda^2})$ separate Eq. (\ref{Dexy}) from (\ref{Dez}), unfolding the asymptotic degeneracy of planar and perpendicular transversal modes at $|\lambda| \rightarrow \infty $.

\section{Boundary Layer}
\label{Section VII}

 In the following we attempt to validate our theory by exploring an isolating mechanism to allow a sensible modelling of nature. Atomic gases containing Avogadro's number of atoms pose a many-body problem where each atom suffers perturbations from all other atoms. Electromagnetic isolation requires globally bounded orbits with vanishing far fields, in order to decouple individual atoms from experimental boundaries and/or other atoms. Circular orbits are \emph{not} good candidates because these create non-vanishing far-fields. In Ref. \cite{Minimizers} it is shown that globally bounded two-body orbits with vanishing far-fields \emph{must} involve velocity discontinuities.
\par
The non-zero real part of growth-rate Eq. (\ref{unperastar}) suggests that there are no other near-circular $C^2$ solutions besides circular orbits, and the linearized modes blow up at either $t \to \pm \infty $. This unless a corner point invalidates the linearization. Therefore, we are led again to seek globally bounded extrema \emph{with corner points}.
\par
\begin{figure}[h!]
   \centering
  \includegraphics[scale=0.5]{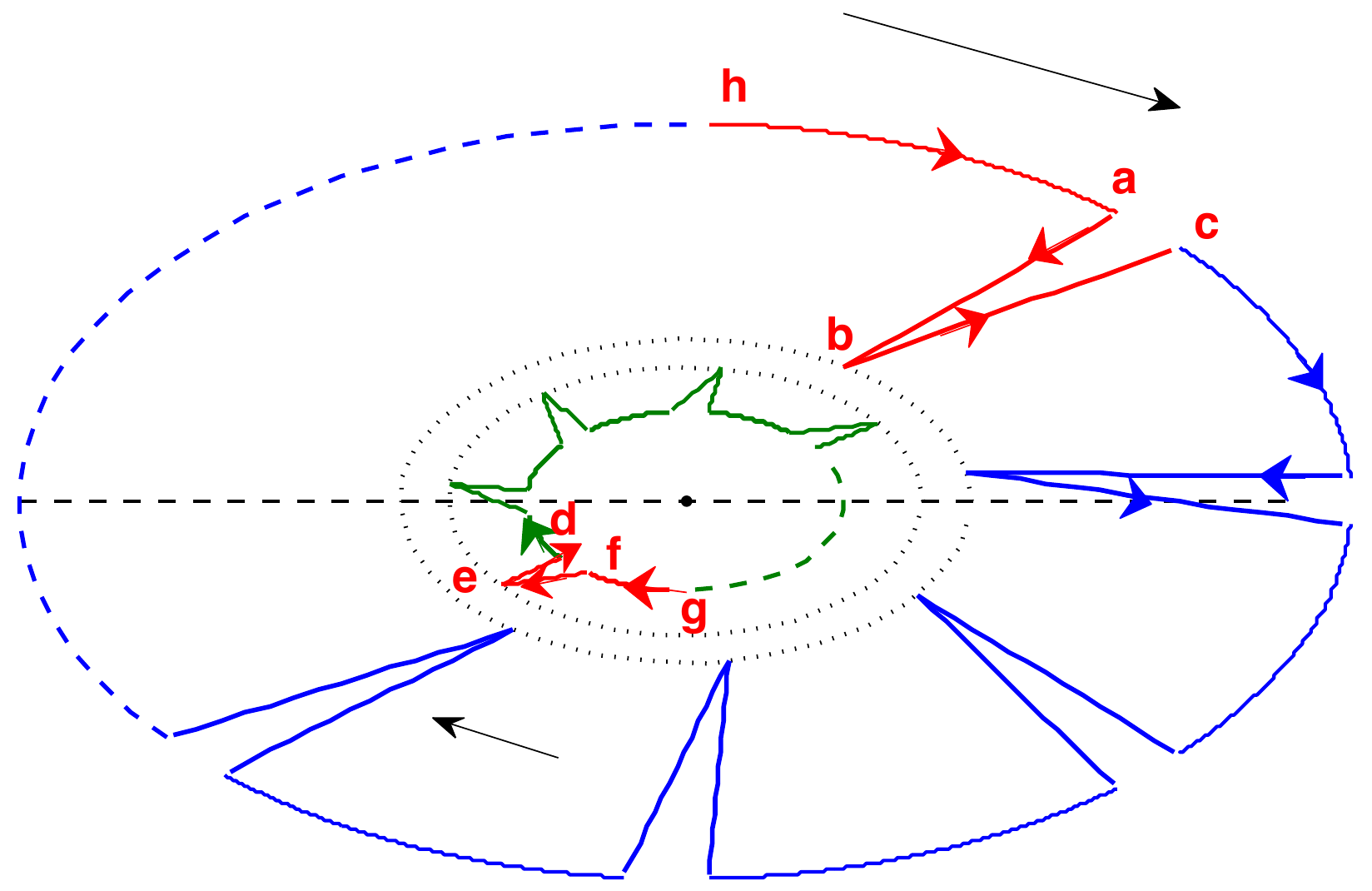} 
   \caption{Illustrated are (i) elsewhere boundary segment (red) of particle 1 with a spike along the advanced light-cone (ab) to corner point (b); and (ii) elsewhere boundary segment 2 with a spike to corner point (e) (red line). At points (a) and (c), perturbed orbit 1 (red and blue solid lines) have a corner to/from the 180-degree corners (b) and (e). Boundary layer magnitudes are exaggerated for illustrative purposes. Arbitrary units.}
  \label{fig3}
\end{figure}

 
Figure \ref{fig3} illustrates a pair of boundary segments containing boundary-layer regions of fast motion along the light-cone separation, segments $abc$ and $fed$, henceforth \emph{spikes}. As explained in Ref. \cite{BellenZennaro}, a neutral differential delay equation (NDDE) can propagate the discontinuity to the next segment (see method of steps and examples of NDDE's versus ODE's in Ref. \cite{BellenZennaro}). In other words, spikes along the orbit are created by spikes inside the boundary segments illustrated in red in FIG. \ref{fig3}.

\par
We set about the task to construct a periodic broken extremum
using a boundary-layer perturbation that assumes regular segments separated by boundary layer regions (spiky segments) each containing one or more corner points, as illustrated in FIG. \ref{fig3}. Along both trajectories, boundary layers have an angular width $\alpha \theta \ll \theta$. Outside boundary layers we can linearize because trajectories are $C^{2}$ and deviating arguments fall on $C^2$ segments as well. Inside boundary layers we \emph{do not} linearize but rather use a variational approximation using head-on collisional trajectory segments, and apply the necessary condition (\ref{WE1}) at rebouncing corners. 
\par
The roots of Eqs. (\ref{Dexy}) and (\ref{Dez}) near
each root (\ref{unperastar}) of the limiting Eq. (\ref{detxy0}) are, respectively, 
\begin{eqnarray}
\lambda _{xy}(\theta,q ) &\equiv &\sigma _{xy}+\pi qi+i\epsilon _{xy}, \label{Lqxy} \\
\lambda _{z}(\theta,q ) &\equiv &\sigma _{z}+\pi qi+i\epsilon _{z},  \label{Lqz}
\end{eqnarray}%
where $q$ is an arbitrary integer, $\epsilon _{xy}(\theta,q )=-\epsilon_{xy}(\theta, -q)$ and $\epsilon _{z}(\theta,q )=-\epsilon_{z}(\theta, -q)$ (by the symplectic symmetry). An  $O(\mu \theta^4/M)$ approximation for positive $\sigma$ is obtained by setting $4\cosh^2 (\lambda)  \approx 2\sinh(2\lambda) \approx\exp(2\lambda)$ into Eqs. (\ref{Dexy}) and (\ref{Dez}), thus defining polynomials
\begin{eqnarray}
F_{xy}(\lambda)&=&1-\frac{\mu \theta^4}{4M}\exp(2\lambda) (1-\frac{2}{\lambda}+\frac{7}{\lambda^2}-\frac{10}{\lambda^3}+\frac{5}{\lambda^4} )\notag \\ &\equiv&1-\frac{\mu \theta^4}{4M}\exp(2\lambda)p_{xy}(\frac{1}{\lambda})=0, \label{polyxy}
\end{eqnarray}
and
\begin{eqnarray}
F_{z}(\lambda)&=&1-\frac{\mu \theta^4}{4M}\exp(2\lambda) (1-\frac{2}{\lambda}-\frac{1}{\lambda^2}+\frac{2}{\lambda^3}+\frac{1}{\lambda^4} )\notag \\ &\equiv&1-\frac{\mu \theta^4}{4M}\exp(2\lambda)p_{z}(\frac{1}{\lambda})=0. \label{polyz}
\end{eqnarray}
It follows from Eqs. (\ref{Dexy}) and (\ref{Lqxy}) that 
\begin{eqnarray}
\exp(4i\epsilon_{xy})=\frac{p_{xy}(\frac{1}{\lambda_{xy}^*}) }{ p_{xy}(\frac{1}{\lambda_{xy}})} \label{exy},
\end{eqnarray}
while Eqs. (\ref{Dez}) and (\ref{Lqz}) yield
\begin{eqnarray}
\exp(4i\epsilon_{z})=\frac{p_{z}(\frac{1}{\lambda_z^*}) }{ p_{z}(\frac{1}{\lambda_z})} \label{ez}.
\end{eqnarray}
\par
Disregarding the contribution of $\epsilon_{xy}$ and $\epsilon_z$ and replacing $\lambda_{xy}= \lambda_z =\sigma + \pi q i$\; into the right-hand-sides of (\ref{exy}) and (\ref{ez})   yields, to the first order in $(1/\lambda)$ 
\begin{eqnarray}
\epsilon_{xy} \simeq \epsilon_z = \Im(\frac{1}{\lambda})=\frac{-\pi q}{ (\sigma^2 + \pi^2 q^2)} . \label{eo1}
\end{eqnarray}
We henceforth use a positive integer $q$, so that the \textit{energetic mismatches} $\epsilon_{xy}$ and $\epsilon_z$ predicted by Eq. (\ref{eo1}) are negative $O(\frac{1}{\lambda})$ numbers. 
\par
Another order can be gained expanding the right-hand-side of (\ref{exy}) in a Taylor series on the deviation $\epsilon_{xy}$ about $\lambda_q=\sigma + \pi q i $, which generates only terms at $O(1/\lambda^3)$, so that up to $O(1/ \lambda^2)$ Eq. (\ref{exy}) yields
\begin{eqnarray}
\epsilon_{xy}(q) = \frac{-\pi q}{ (\sigma^2 + \pi^2 q^2)} +\frac{5\pi q \sigma}{(\sigma^2 + \pi ^2 q^2 )^2}, \label{exyO2}
\end{eqnarray}
$\epsilon_{xy}$ negative and monotonically increasing for $q \geq 6$. Analogously, the right-hand-side of (\ref{ez}) evaluated at $\lambda_q=\sigma + \pi q i $ yields $\epsilon_{z}$ up to $O(1/ \lambda^2)$ 
\begin{eqnarray}
\epsilon_{z}(q) = \frac{-\pi q}{ (\sigma^2 + \pi^2 q^2)} -\frac{3\pi q \sigma}{(\sigma^2 + \pi ^2 q^2 )^2}, \label{ezO2}
\end{eqnarray} 
again negative and monotonically increasing for $q \geq 6$.
\par
Let us assume a corner along orbit $1$ at $t=0$, illustrated by the dashed line in FIG. \ref{fig3}. 
We define the first regular layer $ {\alpha} r_b < t< r_b -\alpha r_b $ as the first and last zeros of the (exponentially increasing and fast oscillating) perturbation inside the first light-cone time zone. Using a linear combination of the symplectic quartet of linearized modes, and (\ref{coordinates}) and (\ref{eigenvector}) with  $l_1 \propto L_1=\frac{\mu \theta}{M}$ and  $u_1 \propto U_1=1$, the orbital perturbation constructed to vanish at layer edges is 
\begin{eqnarray}
\negthickspace \negthickspace  \left(\begin{smallmatrix} \delta x_j \\  \delta y_j    \end{smallmatrix} \right)&=& \notag \\
 &&  \negthickspace \negthickspace \negthickspace \negthickspace \left(\begin{smallmatrix} \frac{\mu \theta A }{M} \\  A \end{smallmatrix} \right)  \frac{\cosh(\frac{\sigma_{xy}t}{r_b})}{\cosh(\frac{\sigma_{xy}}{2})}\sin \left( \frac{[\pi q+\epsilon_{xy}(q)-\theta](t- {\alpha} r_b)}{r_b}  \right). \notag \\  \label{combxy}
\end{eqnarray}
From (a) to (h) the velocity of the perturbed orbits oscillate fast, and at points (a) and (h) the perturbation of trajectory $1$ crosses the unperturbed orbit $1$ (the perturbations are not illustrated in FIG. \ref{fig3}). 
In Eq. (\ref{combxy}), $\sigma_{xy}$ and $\epsilon_{xy}$ are given by (\ref{Lqxy}) and we must have $A < \frac{r_b}{\pi q }$, to avoid a superluminal velocity at layer edges. Along the regular region the phase of the sine function in (\ref{combxy}) advances by almost $\pi q$, and the condition of vanishing at $t=r_b -\alpha r_b $ yields $\alpha = \frac{( \epsilon_{xy}-\theta)}{2 \pi q } $. 
\par
The variational approximation for the boundary-layer motion is illustrated in FIG. \ref{fig3}, with corners along a given orbit all equivalent by a $\theta$-rotation. The central spike starts along trajectory $1$ with a ninety-degree corner, point (a), and respective corner in light-cone along trajectory $2$, point (d). Next is a straight-line segment of a collisional trajectory ($\bar{ab}$) terminated by an almost 180-degree corner ((b) and (e)), then a straight-line climb ($\bar{bc}$) to the last ninety-degree corner ((c) and (f)), resuming motion along the next regular layer. 
\par
Corners are nontrivial solutions of condition (\ref{WE1}), and 180-degree corners, (points (b) and (e) in FIG. \ref{fig3}), are simpler to analyse: for a nontrivial corner, condition (\ref{WE1}) requires the other particle to have a velocity discontinuity at least at one of the light-cones. For simplification, we study resonant minimizers where velocities are discontinuous at both light-cones and further specialized to 180-degree corners that are spikes along the local radial direction $\hat{\mathbf{\rho}}_j$ to each circular orbit, $\hat{\mathbf{\rho}}_{j\pm} \cdot \mathbf{v}_{j\pm }^l  =-\hat{\mathbf{\rho}}_{j\pm} \cdot \mathbf{v}_{j\pm }^r$ . Moreover, we consider only periodic minimizers such that all corners are equivalent by a rotation of $\theta$, i.e., satisfy the discrete-rotation-symmetry  $\mathbf{n}_{j+} \cdot \mathbf{v}_{j+ }^l  =\mathbf{n}_{j-} \cdot \mathbf{v}_{j- }^l$ and $\mathbf{n}_{j+} \cdot \mathbf{v}_{j+ }^r  =\mathbf{n}_{j-} \cdot \mathbf{v}_{j-}^r$. 
\par
For planar motion, condition (\ref{WE1}) yields four equations, one along each Cartesian direction and for $i=1,2$. For the assumed radial spikes the vectorial components of (\ref{WE1}) perpendicular to each $\hat{\mathbf{\rho}}_i$ vanish, yielding a $2 \times 2$ linear homogeneous system for the $\hat{\mathbf{\rho}}_j \cdot \Delta \mathbf{v}_j=2 |\mathbf{v}_j| $,  
\begin{equation}
  \left( \begin{smallmatrix} (E_1 +\frac{1}{rD_2}) & -\frac{\cos{\theta}}{rD_2} \\  -\frac{\cos{\theta}}{rD_1} & ( E_2 +\frac{1}{rD_1})   \end{smallmatrix} \right)
  \left(\begin{smallmatrix}  |\mathbf{v}_1|  \\  | \mathbf{v}_2|   \end{smallmatrix} \right)=0. \label{matriz22} 
\end{equation} 
 The vanishing determinant of a nontrivial solution to Eq. (\ref{matriz22}) in the limit $\cos(\theta) \rightarrow 1$ can be expressed as 
\begin{eqnarray}
Det & \;  = \; & E_1 E_2 + \frac{1}{\bar{r}}(\frac{E_1}{D_1}+\frac{E_2}{D_2})=0, \label{stepping-stone}
\end{eqnarray}
where $D_1 \equiv (1-(\mathbf{n}_1 \cdot \mathbf{v}_1)^2)$ and $D_2 \equiv (1-(\mathbf{n}_2 \cdot \mathbf{v}_2)^2)$ are evaluated at the corner and  $\bar{r}$ is the distance from point (b) to point (e) in FIG. \ref{fig3}. Since the $D_i$ are positive, inspection of (\ref{stepping-stone}) shows that at least one energy must be negative for a nontrivial solution. 
\par
The energies defined by (\ref{WE2}) for radial spikes at large separations are the following functions of time
\begin{eqnarray}
E_1 &=&\frac{m_1}{\sqrt{1-v_1^2}} -\frac{1}{\bar{r}D_2}, \label{ene1} \\
E_2 & =&\frac{m_2}{\sqrt{1-v_2^2}} -\frac{1}{\bar{r}D_1}. \label{ene2}
\end{eqnarray} 
Nonlinear Eqs. (\ref{stepping-stone}), (\ref{ene1}) and (\ref{ene2}) can be solved for $\bar{r}$, yielding
\begin{eqnarray}
\bar{r}^2 &=& \frac{1}{m_1 m_2} \frac{ \sqrt{1-v_1^2} }{D_1}  \frac{\sqrt{1-v_2^2} }{D_2}\notag \\ & =& \frac{1}{m_1 m_2} \frac{ \sqrt{1-v_1^2} }{(1-(\mathbf{n}_1 \cdot \mathbf{v}_1)^2)}  \frac{\sqrt{1-v_2^2} }{(1-(\mathbf{n}_2 \cdot \mathbf{v}_2)^2)}, \label{stepping-stone2}
\end{eqnarray}
 proving that a separation in light-cone $\bar{r}$ of the order of the (large) circular radius $r_b=1/\mu \theta^2$ requires that $|\mathbf{n}_j \cdot \mathbf{v}_j| \rightarrow 1 $ at least for one particle, which in turn requires a \emph{quasi-luminal} velocity. Again, partial energies are not constants of motion and may assume the needed spiky negative values only for a \emph{split second} during the boundary-layer time (the exponential blow-up time is $ \frac{\Omega}{2 \pi} \frac{r_b}{\sigma} \simeq \frac{\theta}{2 \pi \sigma}$ circular periods).
\par 
Assuming $m_2 \gg m_1=1$, and a trajectory with a large separation $r_b= 1/ \mu \theta^2$, if $\bar{r}$ in Eq. (\ref{stepping-stone2}) is to be near the large circular radius $ r_1 \approx r_b= \frac{1}{\mu \theta^2}$ of trajectory $1$, then the $(ab)$ straight-line spiky segment of FIG. \ref{fig3} must be (almost) along the light-cone direction such that $D_1 \equiv (1-(\mathbf{n}_1 \cdot \mathbf{v}_1)^2)$ can be small at the corner and the value of $E_2$ negative (possibly negative only for the split second of the spike). The former conditions are achieved simply by letting the electron have a large radial velocity at point $(b)$ of FIG. \ref{fig3}. 
\par
Notice that (\ref{stepping-stone2}) solves the \emph{fully nonlinear} condition (\ref{WE1}) without any approximation; the 180-degree corners function as \emph{stepping-stones} when one particle reaches the critical quasi-luminal velocity necessary for the formation of corners at-a-far-distance. Corner creation at a far distance is a rebouncing mechanism alternative to charges falling into each other. 
 \par
 Last, we discuss if and how ninety-degree corners satisfy the extremum condition, to complete the justification of the minimizer of FIG. \ref{fig3} ( points (a) and (c) in FIG. \ref{fig3}). We recall that perturbation (\ref{combxy}) naturally blows up after each one-light-cone time, generating synchronized and periodic velocity bursts lasting for the very small boundary-layer times, enough to trigger the spikes. The ratio of the small boundary-layer time to the period is a good estimate of the (small) probability of finding a large velocity. 
\par
 For the most general co-planar corner problem, condition (\ref{WE1}) yields a $4 \times 4$ homogeneous system with linear matrix having diagonal elements equal to $E_i$ plus terms of the order of $1/(\bar{r}D_j)$, while the off-diagonal terms are proportional to $1/(\bar{r}D_i)$, a generic form shared by matrix (\ref{matriz22}). Again, the fully nonlinear necessary condition for a 90-degree turn at large separations is the vanishing of the $4 \times 4$ determinant, which requires the vanishing of one of the $E_i$, which in turn requires a large denominator and a quasi-luminal velocity. A condition analogous to (\ref{stepping-stone2}) results from a fully nonlinear analysis if velocities are to increase along the circular orbit right before corner (a) of FIG. \ref{fig3}. Again, the exponential blow-up of Eq. (\ref{combxy}) eventually reaches that necessary quasi-luminal velocity at layer edge if $A \simeq \frac{r_b}{\pi q}$.

\par
The physical intuition about the spikes of FIG. \ref{fig3} is that the synchronised velocity bursts create large-amplitude electromagnetic fields at the other particle. These fields represent photons carrying momentum along the direction $\mathbf{n}$ of particle separation,  and the mechanism of continuous absorption of momentum from transversal electromagnetic waves is discussed in textbooks \cite{Jackson}.  What is necessary to explain within variational electrodynamics is the mechanism to switch from the circular trajectory to a quasi-luminal head-on collisional trajectory at a discontinuous rate above threshold, a collapse caused by the bursting attractions between particles. At a far-distance the spiky behaviour goes unnoticed because the synchronised longitudinal chase has a vanishing net current and produces weak Biot-Savart fields. 
 \par

Weierstrass-Erdmann conditions were used in Ref. \cite{double-slit} to model \emph{double-slit interference} caused by interaction-at-a-distance with the velocity discontinuities of the bounded  trajectories of material electrons inside the grating. Our Eqs. (\ref{momentum1}) and (\ref{WE2}) are exactly Eqs. (16) and (17) of Ref. \cite{double-slit}, while the above explained criteria provided by the vanishing of the partial energies is the content of Eqs. (19) and (20) of  Ref. \cite{double-slit}.
 \par
 
\section{Finitely measured neighbourhood of broken minimizers}
\label{Section VIII}

The tangent dynamics of circular orbits has an infinite number of  \emph{unstable} transversal modes of \emph{arbitrarily large frequencies}, as seen by the linear growth frequencies (\ref{Lqxy}) and (\ref{Lqz}). This is unlike the classical Kepler problem, whose tangent dynamics has a finite number of frequencies of the order of the orbital frequency \cite{PRL}. Since linearized modes  (\ref{Lqxy}) and (\ref{Lqz}) are unstable, continuation along the $C^2$ segment would blow up, and a velocity discontinuity is needed to break away from the $C^2$ segment. A corner requires stepping-stone condition (\ref{stepping-stone}), and for a small neighbourhood of minimizers with corners to exist, circular orbits with planar and perpendicular perturbations require the resonances studied below.
\par
The perpendicular perturbation in the regular region is constructed analogously to the planar perturbation (\ref{combxy}), using the linearized modes explained in appendix B of Ref. \cite{pre_hydrogen}, yielding
\begin{eqnarray}
\delta z_j&=& 
    B \frac{\cosh(\frac{\sigma_{z}}{r_b}t )}{\cosh(\frac{\sigma_{z}}{2})} \sin \left(\frac{[\pi s+\epsilon_{z}(s) ](t-\alpha r_b)}{r_b} \right), \label{combz}
\end{eqnarray}
where $s$ is any integer (possibly different from $q$) and $B$ is an arbitrary amplitude smaller than $\frac{r_b}{\pi s}$ to avoid a superluminal velocity at layer edge. Again, the amplitude of the transversal perturbation (\ref{combz}) is constructed to vanish along the circular orbit at both edges $t=\alpha r_b$ and $ t=r_b -\alpha r_b$, to keep the property that corners see other corners in light-cone (a boundary-layer-adjusted resonance). 
\par
Since the phase of oscillation $ ( \simeq  \pi s /r_b) $ is fast, a large orbital velocity at layer edge results from a small amplitude $B \simeq \frac{r_b}{\pi s} $, which is illustrated in FIG. \ref{fig3}. Layer edge amplitudes of both types of transversal modes should vanish while their derivatives reach a quasi-luminal velocity. Amplitude (\ref{combxy}) vanishes at $t=r_b-\alpha r_b$ if
\begin{equation}
[\pi q+\epsilon_{xy}(q)-\theta](1- 2{\alpha})=\pi q, \label{Cq}
\end{equation}
other multiples of $\pi$ being excluded because $\epsilon_{xy}(q)$, $\theta$ and $\alpha$ are small. Analogously, amplitude (\ref{combz}) vanishes at $t=r_b-\alpha r_b$ if
\begin{equation}
[\pi s+\epsilon_{z}(s)](1- 2{\alpha})=\pi s, \label{Cs}
\end{equation}
where again other integer multiples of $\pi$ are impossible because $\epsilon_{z}(s)$ and $\alpha$ are small. 
\par
 If Eqs. (\ref{Cq}) and (\ref{Cs}) hold, both edges $t=\alpha r_b$ and $t=r_b-\alpha r_b$ have the \emph{same} quasi-luminal $|\mathbf{v}_j|$ for arbitrary amplitudes near $(A,B)= ( \frac{r_b}{ \pi q} , \frac{r_b}{\pi s})$. For this it is \emph{necessary} that
\par
\begin{equation}
\theta= \frac{q\epsilon _{xy}(q)-s\epsilon _{z}(s)}{q}, \label{resonance}
\end{equation}
as obtained eliminating $\alpha$ from Eqs. (\ref{Cq}) and (\ref{Cs}). If condition (\ref{WE1}) holds at corner (c) of FIG. \ref{fig3}, then by  Eq. (\ref{resonance}) it automatically holds at corner (a) of FIG. \ref{fig3}, which carries on to all other corners by the discrete $\theta$ rotational symmetry. Condition (\ref{resonance}) is also a \emph{probabilistic condition}, i.e., given (\ref{resonance}) a \emph{whole neighbourhood} of orbits with $(A,B)$ near $( \frac{r_b}{\pi q} ,\frac{r_b}{\pi s} ) $ is focused like a caustic into the Kernel of the corner point, thus allowing a \emph{finitely measured} neighbourhood of broken extrema.
 \par
\begin{figure}[h!]
   \centering
  \includegraphics[scale=0.43]{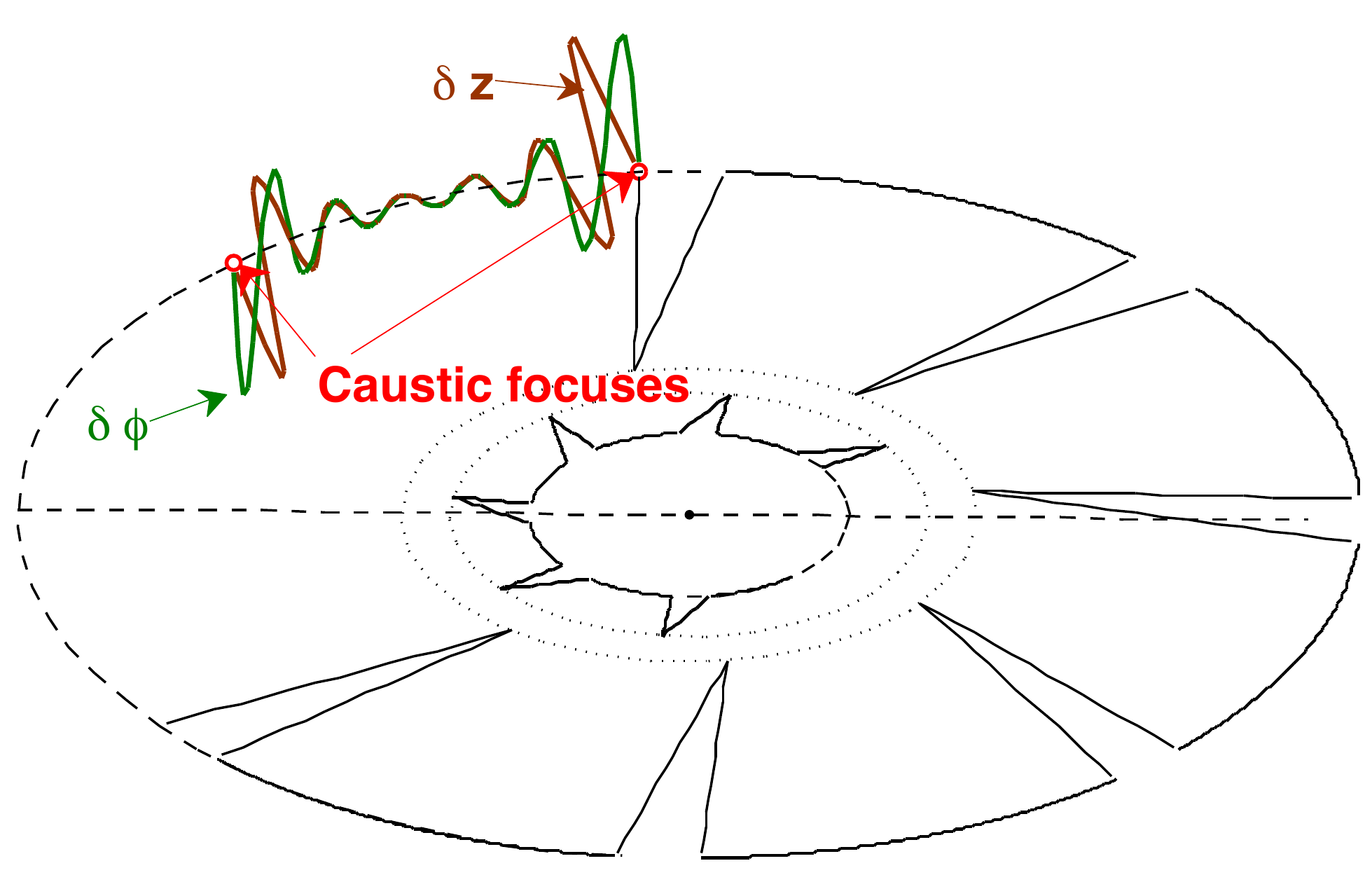} 
\caption{The angular perturbation $\delta \phi$ along the circular orbit is illustrated by the green line. Illustrated is also the transversal perturbation  $\delta z$ (brown). The resonant caustic focus is created when $\delta \phi$ and $\delta z$ perturbations vanish at both ends of each regular segment, by resonance condition (\ref{resonance}). Arbitrary units.  }
  \label{fig5}
\end{figure}

Figure \ref{fig5} illustrates the exponentially exploding transversal perturbations (\ref{combz}) and (\ref{combxy}) being focused into the corners at both layer edges. The perturbations need to be in phase to be focused in and out of both corners with the same large velocity amplitude required by Eq. (\ref{WE1}) for the 90-degree turn at edges, which imposes resonance (\ref{resonance}), henceforth the \emph{external} resonance. Condition (\ref{WE1}) yields a vanishing determinant with a nontrivial null-vector generating a linear space (Kernel). The null vector provides an extra freedom that can be used to make (\ref{WE1}) hold along slightly different orbits, thus generating periodic orbits passing by every point inside a finite volume around the resonant orbit.  

\par
 Next we calculate the magnitudes of the finitely measured minimizers with $q=s$.
For each $q=s$,  condition
(\ref{resonance}) determines a unique
$\theta $ together with Eqs. (\ref{Dexy}) and (\ref%
{Dez}), as listed in Table \ref{hydrotable}.  For comparison, Table \ref{hydrotable} also lists the first line of each
spectroscopic series, i.e., the
circular lines from quantum level $k+1$ to
quantum level $k$. Historically, the series of hydrogen were  named after Lyman, Balmer, Ritz-Paschen, Brackett, etc.
The frequency over reduced mass of the first line of each spectroscopic series in atomic units is the second column of Table \ref{hydrotable}. We used a Newton method in the complex-$\lambda $ plane to solve Eqs. (\ref{Dexy}), (\ref%
{Dez}) and (\ref{resonance}), as used in Ref. \cite{pre_hydrogen} with Dirac's theory. Our calculations for hydrogen used the protonic-to-electronic mass ratio $(m_2/m_1)=1836.1526$. Table \ref{hydrotable} gives frequency over reduced mass calculated by QM for the first line of the $k^{th}$ spectroscopic series (atomic units), numerically calculated orbital frequency over reduced mass (for a suitable integer $q(k)$), $%
(137^{3}\Omega )/\mu =137^{3}\theta ^{2}(\epsilon _{xy}-\epsilon _{z})$, angular momentum of unperturbed circular guide, and integer $q$.
\begin{table}[htp!]
\begin{center}
\begin{tabular}{|l|l|l|l|l|}
\hline
$k$ & $w_{QM}$ & $137^{3}\theta ^{2}(\epsilon _{xy}-\epsilon
_{z}) $ &  $l_{z}=\theta ^{-1}$& $q$ \\ \hline
1 & 3.750$\times $10$^{-1}$  & 3.852$\times $10$^{-1}$& 188.32 &  7 \\ \hline
2 &6.944$\times $10$^{-2}$& 8.919$\times $10$^{-2}$ &   306.67 & 9 \\ \hline
3 &2.430$\times $10$^{-2}$& 2.485$\times $10$^{-2}$ &   469.54 & 11 \\ \hline
4 &1.125$\times $10$^{-2}$  & 1.392$\times $10$^{-2}$ & 569.61& 12 \\ \hline
5 & 6.111$\times $10$^{-3}$ & 8.070$\times $10$^{-3}$ &  683.13& 13 \\ \hline
6 & 3.685$\times $10$^{-3}$  & 4.825$\times $10$^{-3}$ & 810.89& 14 \\ \hline
7 & 2.406$\times $10$^{-3}$  & 2.966$\times $10$^{-3}$ &953.65& 15 \\ \hline
8 & 1.640$\times $10$^{-3}$& 1.870$\times $10$^{-3}$ &  1112.21 & 16 \\ 
\hline
9 &1.173$\times $10$^{-3}$& 1.206$\times $10$^{-3}$ &  1287.27& 17 \\ 
\hline
10 &8.678$\times $10$^{-4}$ & 7.942$\times $10$^{-4}$ &  1479.62 & 18 \\ 
\hline
11 &6.600$\times $10$^{-4}$ & 5.329$\times $10$^{-4}$ &  1690.02 & 19 \\ 
\hline
12 & 5.136$\times $10$^{-4}$ & 3.639$\times $10$^{-4}$ &  1919.20& 20 \\ 
\hline
\end{tabular}
\end{center}
\caption {Numerical calculations for hydrogen with $(m_2/m_1)=1836.1526$. 
Quantum number $k$ of the circular Bohr transition $k+1\rightarrow
k$, frequency over reduced mass of the
circular QM line in atomic units, $w_{QM}\equiv \frac{1}{2}(%
\frac{1}{k^{2}}-\frac{1}{(k+1)^{2}})$, orbital frequency over reduced mass in atomic units, $ (\Omega/ \mu) =137^{3}\theta ^{2}(\epsilon _{xy}-\epsilon _{z})$, angular momentum of unperturbed orbit in units of $%
e^{2}/c$, $l_{z}=\theta ^{-1}$, and integer $q$.} 
\label{hydrotable}
\end{table}

\par
Table \ref{hydrotable} is to be compared with Table I of Ref. \cite{pre_hydrogen}, which discusses Dirac's electrodynamics with  self-interaction. The same surprising agreement is found in Ref. \cite{pre_hydrogen}. Notice that we also skipped the $q=10$ value in Ref. \cite{pre_hydrogen}, again because the resonance condition is only necessary.
\par
As mentioned below Eqs. (\ref{exyO2}) and (\ref{ezO2}), the energetic mismatches are monotonically increasing for $q \geq 6$, and the frequencies of  Table \ref{hydrotable} agree within ten percent with the first twelve circular hydrogen lines starting from $q=7$ at $k=1$, i.e., $q(1)=7$. Since condition (\ref{resonance}) is only \emph{necessary} (and not sufficient), some values of $q$ may correspond to unstable orbits. This is analogous to the description by QM \cite{Bethe}, where there are selection rules on top of \emph{three} conditions involving integer quantum numbers, and Table \ref{hydroextra} includes the lines that were skipped in Table \ref{hydrotable}.  A theory for the $q$'s that were skipped is presently lacking. Inspection of Table \ref{hydroextra} shows that for $q=1,2,3,4,5,6$ condition (\ref{resonance}) predicts $\theta$ still in the atomic range, but the angular momentum spacing is about half of Planck's constant.  Table \ref{hydroextra} also includes the skipped numerical calculations for $q=8$ and $q=10$. 
\par
Analogy with Sommerfeld's quantization suggests there must be \emph{three} conditions like (\ref{resonance}), involving \emph{three} integer quantum indices, reinforcing that our single condition (\ref{resonance}) is only necessary and alone might not determine a stable orbit.
Inspection shows that if the values of Table \ref{hydroextra} were included in Table \ref{hydrotable}, the angular momentum jump from consecutive lines would be much lesser than about a hundred units of $e^2/c$, which is suggestive of what the missing conditions should do.  
 \begin{table}[htp!]
\begin{center}
\begin{tabular}{|l|l|l|}
\hline
 $137^{3}\theta ^{2}(\epsilon _{xy}-\epsilon
_{z}) $ & $l_{z}=\theta ^{-1}$  &  $q$ \\ \hline
1.506 &   119.53 & 1 \\ \hline
 10.653 &  62.27 &  2 \\ \hline
  9.466 & 64.77 &  3 \\ \hline
   4.664 & 82.01& 4 \\ \hline
 2.006 & 108.63  & 5 \\ \hline
   8.622$\times 10^{-1}$ & 143.96 &  6 \\ \hline
    1.808$\times 10^{-1}$ & 242.32 &  8 \\ \hline
     4.6095 $\times 10^{-1}$ & 382.16 &  10 \\ \hline
\end{tabular}
\end{center}
\caption { Numerical calculations for hydrogen with $q< 7$ and $(m_2/m_1)=1836.1526$. 
 Orbital frequency over reduced mass in atomic units, $ (\Omega / \mu) =137^{3}\theta ^{2}(\epsilon _{xy}-\epsilon _{z})$, angular momentum of unperturbed circular guide in units of $e^2/c $, $l_{z}=\theta ^{-1}$, and integer $q$.} 
\label{hydroextra}
\end{table}
\par
 In the days of Bohr, only twelve lines of the Balmer series could be observed with vacuum tubes, and about thirty-three from celestial spectra \cite{Bohr}. Surprizingly, the emission frequencies agree better with QM for the decays from the twelve deepest quantum levels. As explained in \cite{pre_hydrogen}, the cancelation of dipolar far-fields involves quadratic terms that might require larger amplitudes at large $q$. Given that the $xy$ modes modify the unperturbed $z$-angular momentum more and more at larger $q$, our perturbative results should get worse at larger $q$.

\par
In Table \ref{muotable} we give the numerical calculations for muonium using the positive-muon-to-electron mass ratio $(m_2/m_1)=1836.1526/9 $. Table \ref{muotable} lists
the frequency over reduced mass of the first line of each spectroscopic series as calculated with QM (in atomic units), orbital frequency in atomic units, $%
(137^{3}\Omega )/\mu =137^{3}\theta ^{2}(\epsilon _{xy}-\epsilon _{z})$, and angular momentum of unperturbed circular orbit in units of $e^2/c$. The agreement of the numerical calculations with the atomic magnitudes and QM is again within a few percent for frequencies. 
\begin{table}[htp!]
\begin{center}
\begin{tabular}{|l|l|l|l|l|}
\hline
$k$ & $w_{QM}$ & $137^{3}\theta ^{2}(\epsilon _{xy}-\epsilon
_{z}) $ &  $l_{z}=\theta ^{-1}$& $q$  \\ \hline
1 & 3.750$\times $10$^{-1}$ & 4.039$\times $10$^{-1}$ & 185.63 & 7 \\ \hline
2 & 6.944$\times $10$^{-2}$ & 8.762$\times $10$^{-2}$ &  308.94 & 9 \\ \hline
3 & 2.430$\times $10$^{-2}$ & 2.356$\times $10$^{-2}$ &  478.65& 11 \\ \hline
4 & 1.125$\times $10$^{-2}$& 1.304$\times $10$^{-2}$ &  582.96 & 12 \\ \hline
5 & 6.111$\times $10$^{-3}$  & 7.491$\times $10$^{-3}$ & 701.31& 13 \\ \hline
6 &3.685$\times $10$^{-3}$ & 4.445$\times$10$^{-3}$ &  834.53 & 14 \\ \hline
7 &2.406$\times $10$^{-3}$& 2.717$\times $10$^{-3}$ & 983.41 & 15 \\ \hline
8 &1.640$\times $10$^{-3}$& 1.704$\times $10$^{-3}$ &   1148.75 & 16 \\ 
\hline
9 & 1.173$\times $10$^{-3}$ & 1.095$\times$10$^{-3}$ & 1331.34 & 17 \\ 
\hline
10 & 8.678$\times $10$^{-4}$  & 7.187$\times $10$^{-4}$ & 1531.19& 18 \\ 
\hline
11 & 6.600$\times $10$^{-4}$  & 4.810$\times $10$^{-4}$ & 1751.36& 19 \\ 
\hline
12 &  5.136$\times $10$^{-4}$ & 3.277$\times $10$^{-4}$ & 1990.34& 20 \\ 
\hline
\end{tabular}
\end{center}
\caption{ Numerical calculations for muonium with $(m_2/m_1)=1836.1526/9$.
Quantum number $k$ of the circular Bohr transition $k+1\rightarrow
k$, frequency over reduced mass of the
circular QM line in atomic units, $w_{QM}\equiv \frac{1}{2}(%
\frac{1}{k^{2}}-\frac{1}{(k+1)^{2}})$, orbital frequency over reduced mass in atomic units, $ (\Omega/ \mu) =137^{3}\theta ^{2}(\epsilon _{xy}-\epsilon _{z})$, angular momentum of unperturbed circular guide in units of $%
e^{2}/c$, $l_{z}=\theta ^{-1}$, and integer $q$.}
\label{muotable}
\end{table}
\par
Last, Table \ref{positable} gives the numerical calculations for positronium using the mass ratio $(m_2/m_1)=1$: Table \ref{positable} lists the frequency over reduced mass of the first line of each spectroscopic series calculated by QM (in atomic units), orbital frequency over reduced mass in atomic units and  the angular momentum of the unperturbed circular orbit in units of $e^2/c$. 
\par
Notice in Table \ref{positable} that for positronium the values of $l_z =1/ \theta$ are consistently larger. Using again the fact that condition (\ref{resonance}) is only necessary, Table \ref{positable} starts the $q$  when the angular momentum spacing is about constant, i.e., at $q=8$.  For positronium the numerical calculations find the first root $1/ \theta=40.501$ only at $q=3$, \emph{again} in the atomic magnitude.  The spectrum agrees with 
QM within less than a few percent for the circular lines of the first 12 series. 
\par
 In the former three cases, agreement with emission lines for $k>13$ slowly deteriorates, suggesting that the corresponding minimizers are becoming far from planar. A one-to-one comparison with natural spectra should wait the investigation of broken extrema with spikes filling a tridimensional region. 

\begin{table}[htp!]
\begin{center}\begin{tabular}{|l|l|l|l|l|}
\hline
$k$ & $w_{QM}$ & $137^{3}\theta ^{2}(\epsilon _{xy}-\epsilon
_{z}) $ &  $l_{z}=\theta ^{-1}$& $q$  \\ \hline
1 &  3.750$\times $10$^{-1}$ & 3.423$\times $10$^{-1}$ & 246.76& 8 \\ \hline
2 & 6.944$\times $10$^{-2}$ & 7.749$\times $10$^{-2}$ &  404.86& 10 \\ \hline
3 &2.430$\times $10$^{-2}$& 2.189$\times $10$^{-2}$ &  617.037& 12 \\ \hline
4 &1.125$\times $10$^{-2}$ & 1.240$\times $10$^{-2}$ &  745.61 & 13 \\ \hline
5 &6.111$\times $10$^{-3}$ & 7.286$\times $10$^{-3}$ &  890.33 & 14 \\ \hline
6 & 3.685$\times $10$^{-3}$& 4.416$\times $10$^{-3}$ &  1052.06 & 15 \\ \hline
7 & 2.406$\times $10$^{-3}$ & 2.753$\times $10$^{-3}$ &  1231.64& 16 \\ \hline
8 & 1.640$\times $10$^{-3}$& 1.759$\times $10$^{-3}$ & 1429.92& 17 \\ 
\hline
9 & 1.173$\times $10$^{-3}$ & 1.149$\times $10$^{-3}$ & 1647.73 & 18 \\ 
\hline
10 &8.678$\times $10$^{-4}$ & 7.667$\times $10$^{-4}$ &  1885.88 & 19 \\ 
\hline
11 &6.600$\times $10$^{-4}$ & 5.209$\times $10$^{-4}$ &  2145.20 & 20 \\ 
\hline
12 &5.136$\times $10$^{-4}$ & 3.599$\times $10$^{-4}$ &  2426.50 & 21 \\ 
\hline
\end{tabular}
\end{center}
\caption {Numerical calculations for positronium with $(m_2/m_1)=1$. 
Quantum number $k$ of the circular Bohr transition $k+1\rightarrow
k$, frequency over reduced mass of the
circular QM line in atomic units, $w_{QM}\equiv \frac{1}{2}(%
\frac{1}{k^{2}}-\frac{1}{(k+1)^{2}})$, orbital frequency over reduced mass in atomic units, $ (\Omega/ \mu) =137^{3}\theta ^{2}(\epsilon _{xy}-\epsilon _{z})$, angular momentum of unperturbed orbit in units of $%
e^{2}/c$, $l_{z}=\theta ^{-1}$, and integer $q$.}
\label{positable}
\end{table}

\par
 The numerical calculations up to $q=43$ (not shown) reveal an increasing angular momentum spacing at larger $q$'s. The agreement of the numerical calculations with an effective angular momentum separation seems to continue within thirty percent, suggesting that one can approximate a large number of eigenvalues near the discrete spectrum of Schroedinger's equation.

\par
The agreement of the numerical calculations with a universal value for the fine-structure constant is due to the logarithmic dependence of $ \sigma \equiv  \ln (\sqrt{\frac{4M}{\mu \theta ^{4}}})$ on $(\mu/M)$, as  explained above Eq. (\ref{unperastar}). Formulas (\ref{exyO2}) and (\ref{ezO2}) yield
\begin{eqnarray}
\frac{1}{ \theta}=\frac{1}{\epsilon_{xy}-\epsilon_z} =\frac{ (\sigma^2 (\theta) + \pi ^2 q^2)^2}{8\pi q\sigma(\theta) }, \label{approx}
\end{eqnarray}
an implicit equation for $\theta$, with $ \sigma(\theta) \equiv  \ln (\sqrt{\frac{4M}{\mu \theta ^{4}}})$. The roots of  (\ref{approx}) are insensitive to changes in $(\mu / M)$ over \emph{three orders of magnitude}, e.g., for $q=7$ and $(\mu /M) =1/1837$ (hydrogen), Eq. (\ref{approx}) yields $\theta \simeq 190.09$, while for $q=7$ and $(\mu /M)=9/1836$ (muonium), Eq. (\ref{approx}) yields $\theta \simeq 187.17$, and last for $q=7$ and $(\mu /M)=1/4$ (positronium), Eq. (\ref{approx}) yields $\theta \simeq 186.81$. The numerically calculated values of $l_z \equiv \theta^ {-1}$ fall approximately between the consecutive Bohr orbits $k$ and $k+1$. Tables \ref{hydrotable}, \ref{muotable} and \ref{positable} show that frequencies of spectral lines agree even better with the spectroscopic series.
 \par
Notice that the orbital frequencies determined by (\ref{resonance}) are expressed as a difference of two spectroscopic terms, just like the Rydberg-Ritz principle of atomic physics, i.e.,
\begin{equation}
\Omega \equiv \mu \theta^3= \frac{ \mu \theta \epsilon _{xy}(q) - \mu \theta \epsilon _{z}(s)} {(1/\theta)}, \label{Rydberg}
\end{equation}
with spectroscopic terms $\mu \theta \epsilon _{xy}(q)$ and $ \mu \theta \epsilon _{z}(s)$ defined by the eigenvalues of two \emph{linear} and \emph{infinite dimensional} eigenvalue problems, i.e., Eq. (\ref{4eqs}) and Eq. (B16) of Appendix B in Ref. \cite{pre_hydrogen}.

\par

Last, the far-field-vanishing mechanism of Ref. \cite{pre_hydrogen} used the quadratic term of far-field (\ref{electric}) created by the last right-hand-side term of (\ref{defu}), i.e.,
\begin{equation}
 \delta \mathbf{E}^{(2)}_{j\pm} \equiv \pm e_j  \frac{\mathbf{n}_{j\pm}\times (\delta \mathbf{v}_{j\pm} \times \delta \mathbf{a}_{j\pm} )} { r_{j\pm}}. \label{quadratico} 
\end{equation}
 Equation (\ref{resonance}) with $q=s$ is equivalent to the necessary condition for quadratic term (\ref{quadratico}) to cancel the unperturbed dipolar far-fields by a \emph{resonance} between the fast frequencies of modes (\ref{Lqxy}) and (\ref{Lqz}), i.e., Eq. (56)  of Ref. \cite{pre_hydrogen}.

\section{Discussions and conclusion}

\label{Section IX}

Our hydrogenoid model involves large but \emph{finite} denominators. Assuming $\bar{r} \simeq r_b=1/ \mu \theta^2 $, with $\theta$ taken from either Tables \ref{hydrotable}, \ref{muotable} or \ref{positable} and using a protonic velocity much smaller than the near-luminal electronic velocity (i.e., $v_2 =O(\theta) \simeq 0$ and $ |\mathbf{n}_1 \cdot \mathbf{v}_1| \rightarrow |\mathbf{v}_1| \simeq 1$), Eq. (\ref{stepping-stone2}) predicts a finite value for the spiky denominator in each case, $1/\sqrt{1-v_1^2} = M/ \mu \theta ^4 $, without need of any renormalization. 
\par
The physical (and mathematical) appeal of variational electrodynamics comes from postulating the minimization of a \emph{finite} semi-bounded functional \cite{JMP2009}. According to a theorem of Weierstrass, a semi-bounded continuous functional assumes its \emph{absolute minimum} on a compact set \cite{Brezis,Jabri}, thus creating a well-behaved solution for \emph{arbitrary} continuous and piecewise $C^{2}$ boundary data. In the modern theory of partial differential equations (PDE), a price to pay is a compactification that introduces weak solutions \cite{Brezis}, which are the analogues to our trajectories with corners.
 \par
 A motivation for Wheeler-Feynman theory \cite{Whe-Fey} was Sommerfeld's quantization conditions of Hamiltonian mechanics \cite{Hans_relativistic, HansWKB, Sommerfeld}. Wheeler and Feynman's quantization program stalled because of the lack of a Hamiltonian \cite{Mehra}. Wheeler and Feynman could not have known in 1945 that a finite-dimensional Hamiltonian does not exist for the electromagnetic two-body problem, as proved in 1963 with the no-interaction theorem \cite{Currie,Marmo}. In our generalization, partial energies and momenta appear naturally as eigenvalues of the Weiertrass-Erdmann condition (\ref{WE1}), without resort to a Hamiltonian.
\par
Unlike Dirac's electrodynamics \cite{Dirac}, variational electrodynamics is not ruled out by the Aharonov-Bohm effect \cite{Aharonov}. This experimentally observed effect is a complete paradox for Dirac's electrodynamics of point charges \cite{Dirac}. The origin of the paradox is that along $C^{2} $ smooth orbits the electromagnetic equations of motion involve only \emph{derivatives} of the vector potential, i.e., the Euler-Lagrange equation of partial Lagrangian (\ref{partial}) yields (\ref{Lorentz}) with a right-hand-side equal to $ \nabla U_j +\partial \mathbf{A}_j/ \partial t - \mathbf{v}_i \times (\nabla \times \mathbf{A}_j)$. Instead, along \emph{broken extrema}, the vector potential \emph{$\mathbf{A}_j$ itself} appears on Eq. (\ref{preWE1}), determining an interference-at-a-distance just like in QM, as discussed in Ref. \cite{double-slit}. This further indicates that our generalization of Wheeler-Feynman electrodynamics is experimentally sensible \cite{Whe-Fey}.
 
\par
Last and again, we stress a key difference from the classical principle of least action to the variational principle of Section \ref{Section II}. Namely, the principle of least action is a two-point boundary problem \cite{Fox} that can be turned into an initial-value problem by using initial velocities such that both trajectories arrive precisely at the prescribed endpoints. On the contrary, the relativistic boundary-value problem of Section \ref{Section II} is \emph{really} a boundary-value problem in the sense that it can \emph{not} be turned into an initial-value problem for hidden variables that are set at the initial time. As seen in FIG. \ref{Squema}, the elsewhere boundary segment $(L^-_2, L^+_2)$ plays a non-trivial role by interacting with (and shaping) the finite end-segment of trajectory $(b_1, L_2)$. 
\section{Acknowledgements}
Author acknowledges the partial support of a FAPESP regular grant.


\end{document}